\documentclass[aps, prl, amsmath, amssymb, reprint, longbibliography]{revtex4-2}
\usepackage{graphicx}           
\graphicspath{{figures/}}       
\usepackage{dcolumn}            
\usepackage{bm}                 

\usepackage[utf8]{inputenc}
\usepackage[T1]{fontenc}
\usepackage{mathptmx, etoolbox, dcolumn}
\usepackage{tabularray, ragged2e}
\usepackage{braket}
\usepackage{siunitx,  upgreek, multibib}

\usepackage{dcolumn}
\usepackage{bm}
\usepackage{float}
\usepackage{tabularray}
\usepackage{xr}
\usepackage{xr-hyper} 

\makeatletter
\def\@email#1#2{%
	\endgroup
	\patchcmd{\titleblock@produce}
	{\frontmatter@RRAPformat}
	{\frontmatter@RRAPformat{\produce@RRAP{*#1\href{mailto:#2}{#2}}}\frontmatter@RRAPformat}
	{}{}
}%
\makeatother

\begin{document}

\title{Nonlinear Signal Enhancement of Strongly-Coupled Molecules in Pump--Probe Experiments}

\author{Alexander M. McKillop}
\affiliation{Department of Chemistry, Princeton	University, Princeton, New Jersey 08544, USA}

\author{Marissa L. Weichman}
\affiliation{Department of Chemistry, Princeton	University, Princeton, New Jersey 08544, USA}

\date{\today}

\begin{abstract}
Nonlinear spectroscopy is widely used to study the transient dynamics of molecules under strong light-matter coupling, though it remains unclear to what extent uncoupled intracavity molecules obscure signals from the strongly-coupled species of interest.
Pump or probe fields resonant in the strongly-coupled spectral region will preferentially interact with cavity-coupled molecules, but can exhibit severe optical artifacts due to wave interference in the cavity.
On the other hand, non-resonant pump or probe fields having wavelengths at which the cavity mirrors are highly transmissive propagate as traveling waves along the cavity axis, interacting with both coupled and uncoupled intracavity molecules. 
Here, we quantify the contributions of signals from strongly-coupled and uncoupled populations in simulated experiments with different resonant and non-resonant pump--probe configurations.
We find that while resonant schemes maximize selectivity for the signals of strongly-coupled molecules, non-resonant schemes retain surprisingly high sensitivity to these signals while remaining less susceptible to optical artifacts. 

\end{abstract}

\maketitle

\begin{figure}[tbp]
	\centering \includegraphics[width=3in]{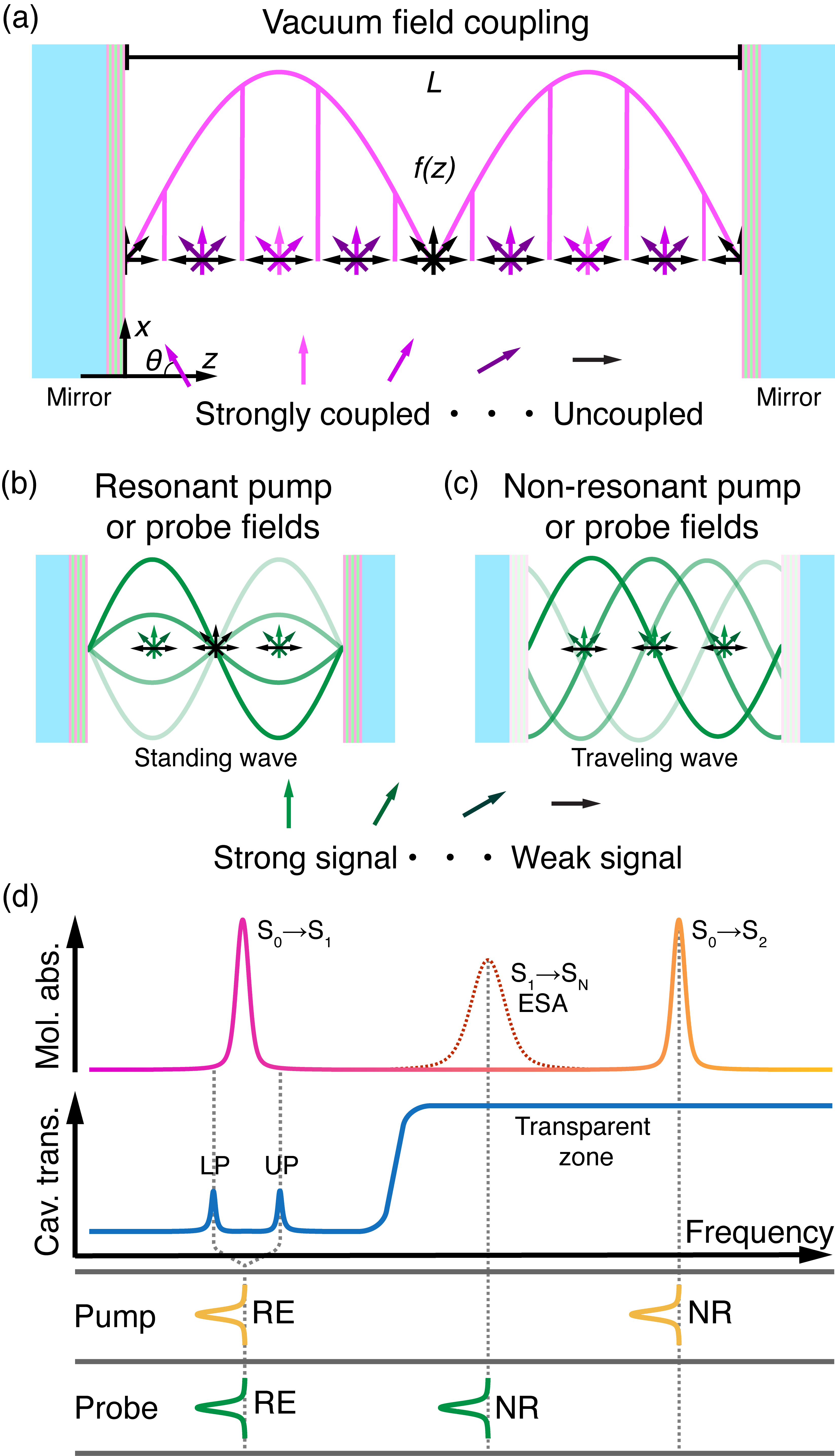}
        \caption{ \label{fig:cavity} 
        (a) The coupling of an intracavity molecule depends on its orientation and position with respect to the cavity field (pink). 
        Molecules that lie near field antinodes or whose transition dipoles are aligned parallel to the cavity field polarization can exchange energy efficiently with the cavity and are thus denoted strongly-coupled (pink vectors) in contrast to uncoupled molecules (black vectors).
        (b) Resonant (RE) and (c) non-resonant (NR) laser fields propagate along the cavity $z$ axis as standing waves and traveling waves, respectively. 
        RE light preferentially interacts with the same molecules whose orientations and $z$ positions also facilitate the strongest cavity-coupling.
        NR light interacts with all molecules along $z$, albeit with a preference for molecules whose transition dipoles are oriented parallel to the laser polarization.
        (d) A schematic of the model system used to illustrate various pump--probe experiments. 
        The top panel illustrates the free-space molecular absorption spectrum, containing the ground-state $\textrm{S}_0\to \textrm{S}_1$ and $\textrm{S}_0\to \textrm{S}_2$ transitions (solid line) and $\textrm{S}_1\to \textrm{S}_\textrm{N}$ excited-state absorption  feature (dashed line). 
        The blue trace shows the cavity transmission spectrum when the $\textrm{S}_0\to \textrm{S}_1$ transition is strongly coupled to form upper and lower polariton peaks (UP, LP). 
        The cavity mirrors are taken to be dichroic, yielding a highly-transmissive ``transparent zone'' at higher frequencies, permitting direct spectroscopic access to the $\textrm{S}_0\to \textrm{S}_2$ and $\textrm{S}_1\to \textrm{S}_\textrm{N}$ bands. 
        Pump (yellow) and probe (green) fields are taken to be RE when their wavelengths fall in the strongly-coupled spectral region and NR when their wavelengths fall in the transparent zone. 
        }
\end{figure}

\textit{Introduction---}
When a cavity mode and an ensemble of intracavity molecules exchange energy faster than their individual decay channels, the system enters the collective strong coupling regime and hybrid-light matter states dubbed polaritons emerge  \cite{ribeiro2018polariton,wang2019strong,herrera2020molecular,simpkins2021mode,dunkelberger2022vibration,bellessa2024materials}.
Polaritons have been studied in the context of photonic control of chemistry \cite{hutchison2012modifying,thomas2016ground,ahn2023modification,lee2024controlling}, 
nonlinear frequency generation \cite{chervy2016high,barachati2018tunable}, lasing \cite{kena2010room,daskalakis2014nonlinear,plumhof2014room},
and other exotic phenomena \cite{dovzhenko2018light,bhuyan2023rise}. 
Within the cavity quantum electrodynamics (cQED) framework, the energetic separation between the two polariton modes is set by the collective Rabi splitting $\hbar\Omega_R$, which scales as $g\sqrt{N}$, where $g$ is the single-molecule coupling strength and $N$ is the number of cavity-coupled molecules \cite{simpkins2015spanning,wright2023rovibrational,mandal2023theoretical}.
Many theoretical treatments invoke the long-wavelength approximation, which assumes that all intracavity molecules share the same $g$.
However, in commonly-used Fabry-P\'erot (FP) cavity geometries, certain molecules are in fact privileged to stronger light-matter coupling than others based on their orientation and position \cite{agranovich2003cavity,litinskaya2004fast,virgili2011ultrafast,canaguier2013thermodynamics,wang2014quantum,shalabney2015enhanced,ahn2018vibrational,delpo2021polariton,mckillop2025cavity}.
In order for a given molecule to participate meaningfully in collective coupling, it must be located near an antinode of the relevant longitudinal cavity mode and its transition dipole must be aligned in the plane of the cavity mirrors \cite{agranovich2003cavity,michetti2005polariton,wang2014quantum,ahn2018vibrational,stemo2022influence,liu2025continuously}.
The intracavity ensemble can therefore be divided into two populations: strongly-coupled (SC) molecules that interact efficiently with the cavity and uncoupled (UC) molecules that lie near field nodes or whose transition dipoles are oriented out of the cavity mirror plane (Fig.~\ref{fig:cavity}a).
Pump--probe spectroscopy is widely used to study ultrafast polariton dynamics and determine whether SC molecules feature distinct behavior compared to the same species in free space \cite{virgili2011ultrafast,schwartz2013polariton,wang2014quantum,xiang2018two,delpo2021polariton,renken2021untargeted,pyles2024revisiting,schwennicke2025molecular,mckillop2026direct}.
The presence of UC molecules may, however, obscure signals from the SC population, depending on the experimental pump--probe configuration.
Many nonlinear polariton spectroscopy experiments use pump or probe light resonant with the spectral region of the polaritons \cite{song2004exciton, schwartz2013polariton,wang2014quantum, avramenko2020quantum, renken2021untargeted, wu2022optical, kushida2025fluidic, rashidi2025efficient}, which we herein refer to as resonant (RE) schemes.
RE light traveling in the cavity at the frequency of a polariton mode will self-interfere to form a standing wave that preferentially interacts with SC molecules at the intracavity field maxima (Fig.~\ref{fig:cavity}b).
However, spectroscopy at RE frequencies is also prone to optical artifacts that arise from spectral filtering and transient changes to the cavity interference conditions \cite{renken2021untargeted,pyles2024revisiting}.
These challenges have inspired many in the field to pivot towards non-resonant (NR) pump or probe schemes to interrogate intracavity systems at wavelengths far removed from the region of strong coupling \cite{liu2021ultrafast,renken2021untargeted,fidler2023ultrafast,chen2024exploring,mckillop2026direct}.
Such implementations often make use of dichroic cavity mirrors that are reflective in one spectral region for strong coupling and highly transmissive in another region \cite{avramenko2020quantum,avramenko2021local,renken2021untargeted,fidler2023ultrafast,chen2024exploring,chen2025ultrafast,mckillop2026direct}. 
NR pump or probe light will pass through transparent cavity mirrors as a traveling wave, interacting indiscriminately with all SC and UC molecules lying along the longitudinal cavity axis (Fig.~\ref{fig:cavity}c).
Pump--probe measurements performed with NR fields may therefore feature fewer cavity artifacts but more significant signal contributions from the UC population.

%
We consider four pump--probe schemes denoted as RE--NR \cite{mckillop2026direct}, NR--NR \cite{fidler2023ultrafast,chen2024exploring,mckillop2026direct}, 
RE--RE \cite{song2004exciton, schwartz2013polariton, wang2014quantum, avramenko2020quantum, renken2021untargeted, wu2022optical, kushida2025fluidic, rashidi2025efficient}, and NR--RE \cite{liu2021ultrafast, renken2021untargeted, chen2025ultrafast}.
To illustrate these schemes more concretely, Fig.~\ref{fig:cavity}d lays out a model system based on the photophysics of a chlorin chromophore recently studied under electronic strong coupling in our lab \cite{mckillop2026direct}.
The top panel of Fig.~\ref{fig:cavity}d plots the free-space molecular absorption spectrum featuring the $\textrm{S}_0\to \textrm{S}_1$ and $\textrm{S}_0\to \textrm{S}_2$ transitions out of the electronic ground state (solid line) and the transient $\textrm{S}_1\to \textrm{S}_\textrm{N}$ excited state absorption (ESA) feature (dashed line) that appears once the $\textrm{S}_1$ state is populated.
As shown in the blue trace in Fig.~\ref{fig:cavity}d, we take the cavity mirrors to be dichroic, featuring high reflectivity at low frequencies to permit strong coupling of the $\textrm{S}_0\to \textrm{S}_1$ transition, and high transmission at higher frequencies to allow NR spectroscopic access to pump the $\textrm{S}_0\to \textrm{S}_2$ transition or probe the $\textrm{S}_1\to \textrm{S}_\textrm{N}$ ESA feature. 
In each pump--probe scheme, we assume that the strongly-coupled $\textrm{S}_1$ manifold is populated via direct RE pumping of $\textrm{S}_0\to\textrm{S}_1$ or via NR pumping of $\textrm{S}_0\to\textrm{S}_2$ followed by rapid internal relaxation to $\textrm{S}_1$ (yellow pulses in bottom of Fig.~\ref{fig:cavity}d). 
The subsequent dynamics of the system may be tracked with an RE probe addressing transient changes in cavity transmission in the strongly-coupled region, or with an NR probe to directly track the ESA feature (green pulses in bottom of Fig.~\ref{fig:cavity}d). 
%
%

%
Here, we use semiclassical simulations to quantify the relative signal amplitudes from intracavity SC and UC molecules in the four pump--probe experimental configurations, accounting for how RE and NR fields travel through the system.
We find, intuitively, that RE fields offer stronger selectivity for exciting and probing SC molecules while NR fields interact more uniformly with both SC and UC molecules.
However, in all pump--probe configurations, the small population of SC molecules contribute more strongly to nonlinear signals than would perhaps be expected.
This over-representation of SC molecules arises because the same orientational factor that renders a dipole strongly coupled also allows it to interact more strongly with the incoming laser fields.
Our simulations therefore indicate that schemes involving an NR pump or probe still retain high sensitivity to the dynamics of SC molecules, alleviating concerns that UC molecules completely obscure the transient response of SC molecules in recent experiments \cite{fidler2023ultrafast,chen2024exploring,mckillop2026direct}.

\begin{figure}[t]
	\centering \includegraphics[width=3in]{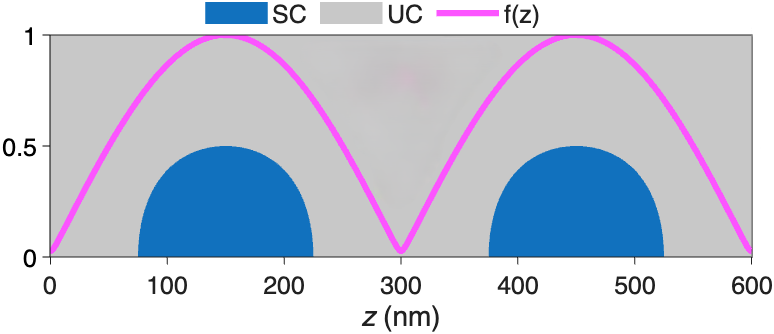}
        \caption{\label{fig:SC} 
        Stacked bar graph of the fractions of strongly-coupled (SC, blue) and uncoupled (UC, gray) molecules along the $z$ axis of an $L=600$ nm cavity, considering resonant coupling to the $m=2$ longitudinal mode whose field profile is plotted in pink. 
        The molecules are assumed to be homogeneously distributed in position and orientation over 2000 $z$ points and 500 $\theta$ points.
        20\% of intracavity molecules meet the threshold for being SC defined in Eq.~\ref{eq:threshold}. 
        }
\end{figure}

\textit{Single-molecule cavity coupling---}
Consider a planar FP cavity of length $L$ filled with a one-dimensional array of non-interacting intracavity molecules along the longitudinal $z$ axis (Fig.~\ref{fig:cavity}a).
We take the $j^\textrm{th}$ molecule to have position $z_j$ and $\textrm{S}_0\to \textrm{S}_1$  transition dipole vector $\vec{\mu}_j$ lying at orientational angle $\theta_j$ with respect to the $z$ axis in the $xz$ plane.
We assume a homogeneous distribution of $z_j$ and $\theta_j$ which will allow us to approximate the intracavity material as a bulk dielectric in treating how laser fields propagate through the intracavity medium (\textit{vide infra}). 
Note that we only consider intracavity fields that are linearly polarized along $x$ and which strike the cavity at normal incidence (e.g., with no in-plane momentum).
We take the coupling strength of the individual $j^\textrm{th}$ molecule to the cavity vacuum field to be given by \cite{mandal2023theoretical}: 
%
\begin{align} \label{eq:g_general}
    g_j = &\vec\mu_j \cdot \vec{E}_\textrm{vac}(\nu_m,z_j)
        = \mu \cdot \sin(\theta_j) \cdot \sqrt{\frac{h\nu_m}{2\epsilon_0 V}} \cdot f(z_j) 
\end{align}
where $\mu$ is the $\textrm{S}_0\to \textrm{S}_1$ transition dipole magnitude, taken to be equivalent for all molecules, 
$\vec{E}_\textrm{vac}(\nu_m,z_j)$ is the electric field amplitude of the $m^\textrm{th}$-order longitudinal cavity mode with frequency $\nu_m$, 
and $V$ is the cavity mode volume.
The function $f(z)$ describes the absolute value of the field profile of the coupled cavity mode (pink trace in Fig.~1a), defined as:
\begin{align}
    f(z) = \big| \sin( m \pi z/L ) \big|.
\end{align}
See the Supplemental Material (SM) for further discussion of this quantity \cite{SM}.
The maximum coupling strength for a dipole located at a field antinode with $f(z_j)=1$ and aligned with the field polarization with $\theta_j = \pi/2$ is given by:
\begin{align}
    g_\textrm{max} = \mu \cdot \sqrt{\frac{h\nu_m}{2\epsilon_0V}}
\end{align}
in agreement with the conventional cQED result within the long-wavelength approximation \cite{mandal2023theoretical,ying2025collective}.
To distinguish SC and UC molecules, we define a threshold single-molecule coupling strength, $g_\textrm{thresh}$, which establishes a molecule as SC if its cavity coupling strength exceeds that of a molecule lying at a field antinode with $\theta=\pi/4$:
\begin{align}
    g_\textrm{thresh} = g_\textrm{max} \cdot \sin(\pi/4).
    \label{eq:threshold}
\end{align}
Figure \ref{fig:SC} illustrates the spatial distribution of molecules that meet this threshold when coupling to the $m=2$ mode of an $L=600$ nm FP cavity.
%
%
Integrating over $z$ and $\theta$, 20$\%$ of intracavity molecules are found to be SC.
%

\begin{figure*}[t]
	\centering \includegraphics[width=6in]{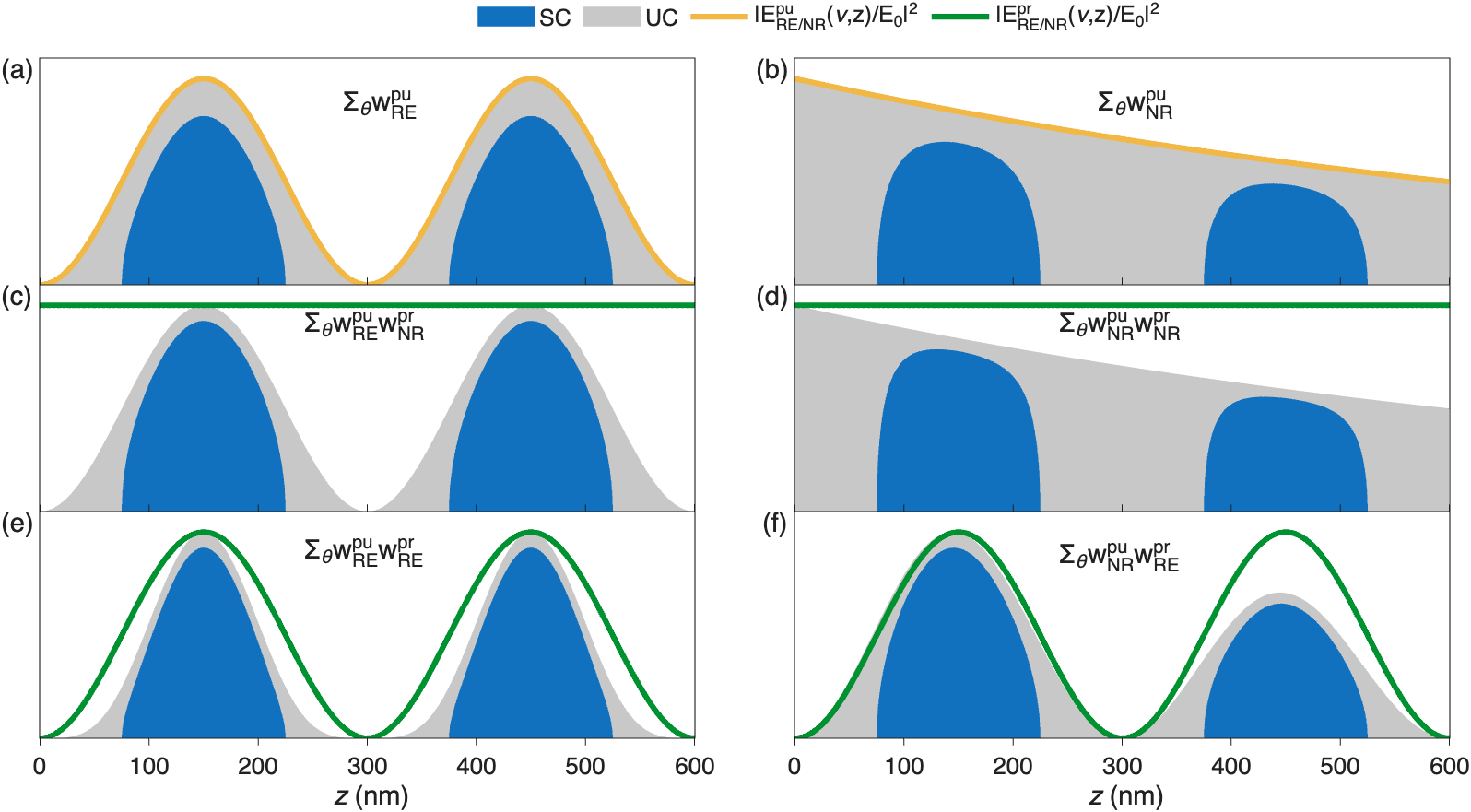}
        \caption{ 
        \label{fig:signals} 
        Relative contributions of strongly coupled (SC, blue) and uncoupled (UC, gray) intracavity molecules to nonlinear pump--probe signals.
        The cavity system is the same as that introduced in Fig.~\ref{fig:SC}.
        (a,b) Stacked bar graphs showing transition weights for the absorption of pump light, $w^\textrm{pu}$, by SC and UC molecules, summed over $\theta$ at each $z$ point for (a) RE and (b) NR pump light.
        Normalized RE and NR pump field intensities are superimposed as yellow lines. 
        (c--f) Stacked bar graphs showing the sum of transition weights for the absorption of both pump and probe light, $w^\textrm{pu} w^\textrm{pr}$, by SC and UC molecules, summed over $\theta$ at each $z$ point for (c) RE--NR, (d) NR--NR, (e) RE--RE and (f) NR--RE pump--probe experiments.
        Normalized NR and RE probe field intensities are superimposed as green lines.
        }
\end{figure*}

\textit{Nonlinear signal enhancement of SC molecules---} 
We now simulate the signal contributions from SC and UC molecules in RE--NR, NR--NR, RE--RE, and NR--RE pump--probe experiments.
For simplicity, we take the $\textrm{S}_0\to \textrm{S}_2$ and $\textrm{S}_1\to \textrm{S}_\textrm{N}$ bands of each molecule to have the same transition dipole magnitude $\mu$ and angle $\theta_j$ as the $\textrm{S}_0\to \textrm{S}_1$ band.
%
%

%
The $j^\textrm{th}$ molecule absorbs light with rate $\Gamma_j$ given by:
\begin{align}
    \Gamma_j \propto \big| \vec{\mu}_j \cdot \vec{E}^\textrm{pu/pr}_\textrm{RE/NR}(\nu, z_j) \big|^2 
     =  \big| \mu \cdot \sin(\theta_j) \cdot E^\textrm{pu/pr}_\textrm{RE/NR}(\nu, z_j) \big|^2
    \label{eq:fgr}
\end{align}
where $E^\textrm{pu/pr}_\textrm{RE/NR}(\nu, z)$ is the electric field amplitude of the pump or probe.
RE fields that are injected into the cavity in the strongly-coupled region form standing waves (Fig.~\ref{fig:cavity}b), yielding an intracavity field intensity $| E^\textrm{pu/pr}_\textrm{RE}(\nu, z) |^2$ that is sinusoidal along $z$. 
NR fields pass through the cavity as traveling waves (Fig.~\ref{fig:cavity}c) with field intensity $| E^\textrm{pu/pr}_\textrm{NR}(\nu, z) |^2$ decaying along $z$ according to the Beer-Lambert Law.
More details are provided on our representation of RE and NR fields in the SM \cite{SM}.
From Eq.~\ref{eq:fgr}, we define a unitless ``transition weight'' which captures the relative propensity of the $j^\textrm{th}$ molecule to absorb pump or probe light:
\begin{align} \label{eq:weights}
    w_j^\textrm{pu/pr} \equiv \big| \sin(\theta_{j}) \big|^2 \; \cdot \; \big| E^\textrm{pu/pr}(\nu, z_j)/E_0 \big|^2
\end{align}
where $E_0$ serves as a normalization factor to account for the field intensity initially incident on the cavity.
Figures~\ref{fig:signals}ab 
plot $w_j^\textrm{pu}$ summed over all values of $\theta$ at each $z$ point for RE and NR pump fields.  
The contributions of SC (blue) and UC (gray) molecules to $w_j^\textrm{pu}$ are shown as stacked bar graphs at each value of $z$, using the same definitions of the SC and UC populations shown in Fig.~\ref{fig:SC}.
As expected, the propensity of a molecule to absorb pump light is strongly dependent on its position in the cavity, particularly for RE pump light.
Figures \ref{fig:signals}cdef illustrate how SC and UC molecules contribute to RE--NR, NR--NR, RE--RE, and NR--RE experiments. 
%
%
An NR probe (Figs.~\ref{fig:signals}cd) only slightly amplifies the contributions of SC molecules due to their alignment with the laser field polarization in the plane of the cavity mirrors, evidenced by comparing the slight increases in blue area between Figs.~\ref{fig:signals}a and \ref{fig:signals}c and between Figs.~\ref{fig:signals}b and \ref{fig:signals}d. 
An RE probe, on the other hand, clearly favors interaction with SC molecules---as evidenced by comparing Fig.~\ref{fig:signals}a to \ref{fig:signals}e and Fig.~\ref{fig:signals}b to \ref{fig:signals}f---because SC molecules are both aligned with the probe field polarization and located at RE field antinodes. 
We can calculate the fractional contribution of SC molecules in each pump-probe configuration:
\begin{align} 
    \label{eq:SC_percent}    
    SC_\textrm{sig} = \sum\limits_{j}^\textrm{SC}w_{j}^\textrm{pu}w_{j}^\textrm{pr} \;\; / \;\; \sum\limits_{j}^\textrm{SC, UC}w_{j}^\textrm{pu}w_{j}^\textrm{pr}
\end{align}
where the sum in the numerator runs only over SC molecules and the sum in the denominator runs over all intracavity molecules. 
Table \ref{tab:table1} lays out $SC_\textrm{sig}$ for the pump--probe configurations shown in Figs.~\ref{fig:signals}cdef,
as well as the ratio of $SC_\textrm{sig}$ to the fraction of molecules designated SC (20\%, per Fig.~\ref{fig:SC}).
While the current model only considers transition dipoles oriented in the $xz$ plane, the same $SC_{sig}$ is obtained when the system is rotated about the $z$ axis, suggesting that this treatment should also hold quantitatively valid for molecules oriented in three dimensions.

It is clear that an RE--RE experiment provides the best sensitivity to signal from SC molecules. 
Surprisingly, though, experiments involving NR fields still provide a significant enhancement of SC signals, with over 2/3 of the transient signal arising from SC species in NR--RE and RE--NR experiments and NR--NR experiments still doubling the contributions of SC molecules compared to the fraction of the population they make up.
We note that the magnitude of the SC signal contribution is determined by our somewhat arbitrary choice of $g_\textrm{thresh}$ in Eq.~\ref{eq:threshold}.
However, we repeat our calculations for additional values of $g_\textrm{thresh}$ in SM \cite{SM} and find that the enhancement of the SC signal persists regardless of the choice of $g_\textrm{thresh}$. 
We observe that as $g_\textrm{thresh}$ increases, the fraction of molecules designated SC decreases, while the SC signal enhancement increases.
Once again, the same orientational factor which renders a molecule strongly coupled also makes it contribute disproportionately to nonlinear signals.
%
%

\begin{figure}[t]
	\centering \includegraphics[width=3in]{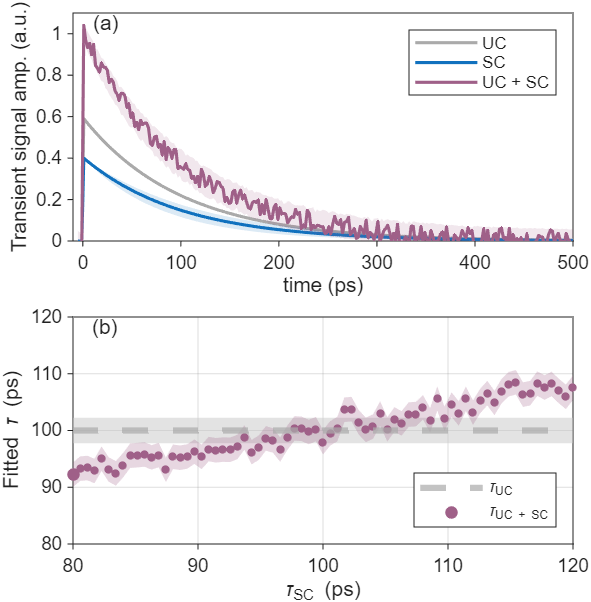}
        \caption{\label{fig:fitting} 
        Model NR--NR pump--probe experiment sensitive to both UC and SC transient signals.
        (a) Simulated decay of transient signals arising from UC molecules with $\tau_\textrm{UC}=100$ ps (gray) and SC molecules with $\tau_\textrm{SC}=80$ to $120$ ps ($100$ ps in blue solid line, $80$ to $120$ ps in pale blue shaded area). 
        Initial signal amplitudes are scaled by their relative contributions in an NR--NR experiment, per Table~\ref{tab:table1}. 
        The combined UC $+$ SC signal for $\tau_\textrm{UC}=100$ ps and $\tau_\textrm{SC}=100$ ps with added random noise whose amplitude is $5\%$ that of the maximum signal amplitude is plotted with a solid purple line. 
        Combined UC $+$ SC traces calculated with $\tau_\textrm{SC}=80$ to $120$ ps cover the pale purple shaded area.
        (b) Time constants from single exponential fitting of the UC $+$ SC traces in panel (a) as a function of $\tau_\textrm{SC}$ (purple dots). 
        The gray dashed line represents $\tau_\textrm{UC} = \SI{100}{ps}$. 
        Shaded areas represent $95\%$ confidence intervals.
        }
\end{figure}

\textit{Experimental sensitivity to cavity-modified dynamics---} 
We finally consider how well one can differentiate signals from SC and UC molecules if the SC population features genuinely distinct cavity-modified dynamics.
We focus on an NR--NR pump--probe configuration in order to illustrate the worst-case scenario for sensitivity to the SC population.
We take the NR probe to read out the $\textrm{S}_1$ lifetime via the $\textrm{S}_1\to \textrm{S}_\textrm{N}$ ESA feature with contributions from both UC and SC molecules.
We take the ESA amplitude to decay as a single exponential with time constant $\tau_\textrm{UC} = \SI{100}{ps}$ for the UC population, and with various cavity-altered time constants for the SC population spanning $\tau_\textrm{SC} = \SIrange{80}{120}{ps}$.
Figure \ref{fig:fitting}a plots the decay of the ESA feature amplitude for UC (gray) and SC (blue) species with the initial signal amplitude scaled by their relative $SC_\textrm{sig}$ contributions in an NR--NR experiment (see Table~\ref{tab:table1}). 
The combined UC $+$ SC traces (purple) represent the total signal arising from both UC and SC molecules, with the addition of random noise whose amplitude is $5\%$ of the maximum signal amplitude. 
%

\begin{table}
        \caption{\label{tab:table1}\ Fractional contribution of SC molecules to pump--probe signals, and signal enhancement factors relative to the fraction of SC molecules ($20\%)$ in different experimental configurations. 
        }
	\begin{tblr}{ colspec={ c  c  c  c c }, rowsep = 3pt }
		\hline\hline
		   & RE--NR & NR--NR & RE--RE & NR--RE \\ \hline
		$SC_\textrm{sig} (\%)$ & $69$ & $40$ &  $79$ & $68$ \\ \hline
	   enhancement & $3.5$ & $2.0$ & $4.0$ & $3.4$ 
            \\ \hline\hline
	\end{tblr}
\end{table}


%
%
Figure \ref{fig:fitting}b plots the results of nonlinear least squares fitting of the combined UC $+$ SC traces to a single exponential decay for various values of $\tau_\textrm{SC}$ (purple points) as $\tau_\textrm{UC}$ is held fixed at \SI{100}{ps} (gray dashed line). 
%
%
When $\tau_\textrm{SC}$ differs from $\tau_\textrm{UC}$ by at least $10\%$, the fitted lifetime differs detectably from $\tau_\textrm{UC}$. 
This finding suggests that, for reasonable experimental noise levels, cavity-modification of the dynamics of SC species by at least $10\%$ from the free-space value should be detectable, even with an NR--NR pump--probe readout and a naive single-exponential fit.
A more appropriate multi-exponential fit could then be subsequently be employed to better disentangle the overlapping SC and UC dynamics.
Further, this example illustrates that if an NR--NR pump–probe experiment returns a null result -- e.g., with UC $+$ SC intracavity dynamics consistent with a free space measurement -- such a measurement can meaningfully constrain the magnitude of any cavity-induced changes in dynamics for SC species.

\textit{Conclusions---} 
We have quantitatively illustrated how transition dipole orientation and position in nonlinear experiments affect the detection of strongly cavity-coupled species using a model system inspired by our own recent experiments \cite{mckillop2026direct}.
Importantly, our results address the concern that UC molecules dominate intracavity pump--probe signals.
We report a significant enhancement of transient signals from SC molecules over UC molecules, implying experimental sensitivity to small changes in SC dynamics.
Ultimately, we show that pump--probe experiments using pulses non-resonant with the strongly-coupled spectral region still retain high sensitivity to the dynamics of strongly-coupled molecules while benefiting from reduced spectral artifacts.
Put another way, these findings suggest that if non-resonant intracavity experiments return transient dynamics consistent with those of the free-space molecular system, then strong statements can indeed be made about a lack of cavity-modified behavior on the part of the SC molecules.
Various additional considerations may be included in future work, including:
treatment of molecular coupling to cavity modes with finite in-plane momentum,
treatment of parallel vs.\ perpendicular pump and probe field polarizations,
and time-evolution of molecular orientation in liquid- or gas-phase intracavity media.

\textit{Acknowledgments---} 
This research was supported by the U.S. Department of Energy, Office of Science, Office of Basic Energy Sciences, CPIMS program, under Early Career Research Program award DE-SC0022948.
This report was prepared as an account of work sponsored by an agency of the United States Government. Neither the United States Government nor any agency thereof, nor any of their employees, makes any warranty, express or implied, or assumes any legal liability or responsibility for the accuracy, completeness, or usefulness of any information, apparatus, product, or process disclosed, or represents that its use would not infringe privately owned rights. Reference herein to any specific commercial product, process, or service by trade name, trademark, manufacturer, or otherwise does not necessarily constitute or imply its endorsement, recommendation, or favoring by the United States Government or any agency thereof. The views and opinions of authors expressed herein do not necessarily state or reflect those of the United States Government or any agency thereof.

\textit{Data availability---} The data that support the findings of this study are available upon request from the corresponding author.

\bibliographystyle{apsrev4-2}
\bibliography{citations}

\providecommand{\noopsort}[1]{}\providecommand{\singleletter}[1]{#1}%
\begin{thebibliography}{50}%
\makeatletter
\providecommand \@ifxundefined [1]{%
 \@ifx{#1\undefined}
}%
\providecommand \@ifnum [1]{%
 \ifnum #1\expandafter \@firstoftwo
 \else \expandafter \@secondoftwo
 \fi
}%
\providecommand \@ifx [1]{%
 \ifx #1\expandafter \@firstoftwo
 \else \expandafter \@secondoftwo
 \fi
}%
\providecommand \natexlab [1]{#1}%
\providecommand \enquote  [1]{``#1''}%
\providecommand \bibnamefont  [1]{#1}%
\providecommand \bibfnamefont [1]{#1}%
\providecommand \citenamefont [1]{#1}%
\providecommand \href@noop [0]{\@secondoftwo}%
\providecommand \href [0]{\begingroup \@sanitize@url \@href}%
\providecommand \@href[1]{\@@startlink{#1}\@@href}%
\providecommand \@@href[1]{\endgroup#1\@@endlink}%
\providecommand \@sanitize@url [0]{\catcode `\\12\catcode `\$12\catcode
  `\&12\catcode `\#12\catcode `\^12\catcode `\_12\catcode `\%12\relax}%
\providecommand \@@startlink[1]{}%
\providecommand \@@endlink[0]{}%
\providecommand \url  [0]{\begingroup\@sanitize@url \@url }%
\providecommand \@url [1]{\endgroup\@href {#1}{\urlprefix }}%
\providecommand \urlprefix  [0]{URL }%
\providecommand \Eprint [0]{\href }%
\providecommand \doibase [0]{https://doi.org/}%
\providecommand \selectlanguage [0]{\@gobble}%
\providecommand \bibinfo  [0]{\@secondoftwo}%
\providecommand \bibfield  [0]{\@secondoftwo}%
\providecommand \translation [1]{[#1]}%
\providecommand \BibitemOpen [0]{}%
\providecommand \bibitemStop [0]{}%
\providecommand \bibitemNoStop [0]{.\EOS\space}%
\providecommand \EOS [0]{\spacefactor3000\relax}%
\providecommand \BibitemShut  [1]{\csname bibitem#1\endcsname}%
\let\auto@bib@innerbib\@empty
\bibitem [{\citenamefont {Ribeiro}\ \emph {et~al.}(2018)\citenamefont
  {Ribeiro}, \citenamefont {Mart{\'\i}nez-Mart{\'\i}nez}, \citenamefont {Du},
  \citenamefont {Campos-Gonzalez-Angulo},\ and\ \citenamefont
  {Yuen-Zhou}}]{ribeiro2018polariton}%
  \BibitemOpen
  \bibfield  {author} {\bibinfo {author} {\bibfnamefont {R.~F.}\ \bibnamefont
  {Ribeiro}}, \bibinfo {author} {\bibfnamefont {L.~A.}\ \bibnamefont
  {Mart{\'\i}nez-Mart{\'\i}nez}}, \bibinfo {author} {\bibfnamefont
  {M.}~\bibnamefont {Du}}, \bibinfo {author} {\bibfnamefont {J.}~\bibnamefont
  {Campos-Gonzalez-Angulo}},\ and\ \bibinfo {author} {\bibfnamefont
  {J.}~\bibnamefont {Yuen-Zhou}},\ }\href@noop {} {\bibfield  {journal}
  {\bibinfo  {journal} {Chem.\ Sci.}\ }\textbf {\bibinfo {volume} {9}},\
  \bibinfo {pages} {6325} (\bibinfo {year} {2018})}\BibitemShut {NoStop}%
\bibitem [{\citenamefont {Hertzog}\ \emph {et~al.}(2019)\citenamefont
  {Hertzog}, \citenamefont {Wang}, \citenamefont {Mony},\ and\ \citenamefont
  {B{\"o}rjesson}}]{wang2019strong}%
  \BibitemOpen
  \bibfield  {author} {\bibinfo {author} {\bibfnamefont {M.}~\bibnamefont
  {Hertzog}}, \bibinfo {author} {\bibfnamefont {M.}~\bibnamefont {Wang}},
  \bibinfo {author} {\bibfnamefont {J.}~\bibnamefont {Mony}},\ and\ \bibinfo
  {author} {\bibfnamefont {K.}~\bibnamefont {B{\"o}rjesson}},\ }\href@noop {}
  {\bibfield  {journal} {\bibinfo  {journal} {Chem.\ Soc.\ Rev.}\ }\textbf
  {\bibinfo {volume} {48}},\ \bibinfo {pages} {937} (\bibinfo {year}
  {2019})}\BibitemShut {NoStop}%
\bibitem [{\citenamefont {Herrera}\ and\ \citenamefont
  {Owrutsky}(2020)}]{herrera2020molecular}%
  \BibitemOpen
  \bibfield  {author} {\bibinfo {author} {\bibfnamefont {F.}~\bibnamefont
  {Herrera}}\ and\ \bibinfo {author} {\bibfnamefont {J.}~\bibnamefont
  {Owrutsky}},\ }\href@noop {} {\bibfield  {journal} {\bibinfo  {journal} {J.\
  Chem.\ Phys.}\ }\textbf {\bibinfo {volume} {152}},\ \bibinfo {pages} {100902}
  (\bibinfo {year} {2020})}\BibitemShut {NoStop}%
\bibitem [{\citenamefont {Simpkins}\ \emph {et~al.}(2021)\citenamefont
  {Simpkins}, \citenamefont {Dunkelberger},\ and\ \citenamefont
  {Owrutsky}}]{simpkins2021mode}%
  \BibitemOpen
  \bibfield  {author} {\bibinfo {author} {\bibfnamefont {B.~S.}\ \bibnamefont
  {Simpkins}}, \bibinfo {author} {\bibfnamefont {A.~D.}\ \bibnamefont
  {Dunkelberger}},\ and\ \bibinfo {author} {\bibfnamefont {J.~C.}\ \bibnamefont
  {Owrutsky}},\ }\href@noop {} {\bibfield  {journal} {\bibinfo  {journal} {J.\
  Phys.\ Chem.\ C}\ }\textbf {\bibinfo {volume} {125}},\ \bibinfo {pages}
  {19081} (\bibinfo {year} {2021})}\BibitemShut {NoStop}%
\bibitem [{\citenamefont {Dunkelberger}\ \emph {et~al.}(2022)\citenamefont
  {Dunkelberger}, \citenamefont {Simpkins}, \citenamefont {Vurgaftman},\ and\
  \citenamefont {Owrutsky}}]{dunkelberger2022vibration}%
  \BibitemOpen
  \bibfield  {author} {\bibinfo {author} {\bibfnamefont {A.~D.}\ \bibnamefont
  {Dunkelberger}}, \bibinfo {author} {\bibfnamefont {B.~S.}\ \bibnamefont
  {Simpkins}}, \bibinfo {author} {\bibfnamefont {I.}~\bibnamefont
  {Vurgaftman}},\ and\ \bibinfo {author} {\bibfnamefont {J.~C.}\ \bibnamefont
  {Owrutsky}},\ }\href@noop {} {\bibfield  {journal} {\bibinfo  {journal}
  {Annu.\ Rev.\ Phys.\ Chem.}\ }\textbf {\bibinfo {volume} {73}},\ \bibinfo
  {pages} {429} (\bibinfo {year} {2022})}\BibitemShut {NoStop}%
\bibitem [{\citenamefont {Bellessa}\ \emph {et~al.}(2024)\citenamefont
  {Bellessa}, \citenamefont {Bloch}, \citenamefont {Deleporte}, \citenamefont
  {Menon}, \citenamefont {Nguyen}, \citenamefont {Ohadi}, \citenamefont
  {Ravets},\ and\ \citenamefont {Boulier}}]{bellessa2024materials}%
  \BibitemOpen
  \bibfield  {author} {\bibinfo {author} {\bibfnamefont {J.}~\bibnamefont
  {Bellessa}}, \bibinfo {author} {\bibfnamefont {J.}~\bibnamefont {Bloch}},
  \bibinfo {author} {\bibfnamefont {E.}~\bibnamefont {Deleporte}}, \bibinfo
  {author} {\bibfnamefont {V.}~\bibnamefont {Menon}}, \bibinfo {author}
  {\bibfnamefont {H.~S.}\ \bibnamefont {Nguyen}}, \bibinfo {author}
  {\bibfnamefont {H.}~\bibnamefont {Ohadi}}, \bibinfo {author} {\bibfnamefont
  {S.}~\bibnamefont {Ravets}},\ and\ \bibinfo {author} {\bibfnamefont
  {T.}~\bibnamefont {Boulier}},\ }\href@noop {} {\bibfield  {journal} {\bibinfo
   {journal} {MRS Bulletin}\ }\textbf {\bibinfo {volume} {49}},\ \bibinfo
  {pages} {932} (\bibinfo {year} {2024})}\BibitemShut {NoStop}%
\bibitem [{\citenamefont {Hutchison}\ \emph {et~al.}(2012)\citenamefont
  {Hutchison}, \citenamefont {Schwartz}, \citenamefont {Genet}, \citenamefont
  {Devaux},\ and\ \citenamefont {Ebbesen}}]{hutchison2012modifying}%
  \BibitemOpen
  \bibfield  {author} {\bibinfo {author} {\bibfnamefont {J.~A.}\ \bibnamefont
  {Hutchison}}, \bibinfo {author} {\bibfnamefont {T.}~\bibnamefont {Schwartz}},
  \bibinfo {author} {\bibfnamefont {C.}~\bibnamefont {Genet}}, \bibinfo
  {author} {\bibfnamefont {E.}~\bibnamefont {Devaux}},\ and\ \bibinfo {author}
  {\bibfnamefont {T.~W.}\ \bibnamefont {Ebbesen}},\ }\href@noop {} {\bibfield
  {journal} {\bibinfo  {journal} {Angew.\ Chem.\ Int.\ Ed.}\ }\textbf {\bibinfo
  {volume} {51}},\ \bibinfo {pages} {1592} (\bibinfo {year}
  {2012})}\BibitemShut {NoStop}%
\bibitem [{\citenamefont {Thomas}\ \emph {et~al.}(2016)\citenamefont {Thomas},
  \citenamefont {George}, \citenamefont {Shalabney}, \citenamefont {Dryzhakov},
  \citenamefont {Varma}, \citenamefont {Moran}, \citenamefont {Chervy},
  \citenamefont {Zhong}, \citenamefont {Devaux}, \citenamefont {Genet} \emph
  {et~al.}}]{thomas2016ground}%
  \BibitemOpen
  \bibfield  {author} {\bibinfo {author} {\bibfnamefont {A.}~\bibnamefont
  {Thomas}}, \bibinfo {author} {\bibfnamefont {J.}~\bibnamefont {George}},
  \bibinfo {author} {\bibfnamefont {A.}~\bibnamefont {Shalabney}}, \bibinfo
  {author} {\bibfnamefont {M.}~\bibnamefont {Dryzhakov}}, \bibinfo {author}
  {\bibfnamefont {S.~J.}\ \bibnamefont {Varma}}, \bibinfo {author}
  {\bibfnamefont {J.}~\bibnamefont {Moran}}, \bibinfo {author} {\bibfnamefont
  {T.}~\bibnamefont {Chervy}}, \bibinfo {author} {\bibfnamefont
  {X.}~\bibnamefont {Zhong}}, \bibinfo {author} {\bibfnamefont
  {E.}~\bibnamefont {Devaux}}, \bibinfo {author} {\bibfnamefont
  {C.}~\bibnamefont {Genet}}, \emph {et~al.},\ }\href@noop {} {\bibfield
  {journal} {\bibinfo  {journal} {Angew.\ Chem.\ Int.\ Ed.}\ }\textbf {\bibinfo
  {volume} {55}},\ \bibinfo {pages} {11462} (\bibinfo {year}
  {2016})}\BibitemShut {NoStop}%
\bibitem [{\citenamefont {Ahn}\ \emph {et~al.}(2023)\citenamefont {Ahn},
  \citenamefont {Triana}, \citenamefont {Recabal}, \citenamefont {Herrera},\
  and\ \citenamefont {Simpkins}}]{ahn2023modification}%
  \BibitemOpen
  \bibfield  {author} {\bibinfo {author} {\bibfnamefont {W.}~\bibnamefont
  {Ahn}}, \bibinfo {author} {\bibfnamefont {J.~F.}\ \bibnamefont {Triana}},
  \bibinfo {author} {\bibfnamefont {F.}~\bibnamefont {Recabal}}, \bibinfo
  {author} {\bibfnamefont {F.}~\bibnamefont {Herrera}},\ and\ \bibinfo {author}
  {\bibfnamefont {B.~S.}\ \bibnamefont {Simpkins}},\ }\href@noop {} {\bibfield
  {journal} {\bibinfo  {journal} {Science}\ }\textbf {\bibinfo {volume}
  {380}},\ \bibinfo {pages} {1165} (\bibinfo {year} {2023})}\BibitemShut
  {NoStop}%
\bibitem [{\citenamefont {Lee}\ \emph {et~al.}(2024)\citenamefont {Lee},
  \citenamefont {Melton}, \citenamefont {Xu},\ and\ \citenamefont
  {Delor}}]{lee2024controlling}%
  \BibitemOpen
  \bibfield  {author} {\bibinfo {author} {\bibfnamefont {I.}~\bibnamefont
  {Lee}}, \bibinfo {author} {\bibfnamefont {S.~R.}\ \bibnamefont {Melton}},
  \bibinfo {author} {\bibfnamefont {D.}~\bibnamefont {Xu}},\ and\ \bibinfo
  {author} {\bibfnamefont {M.}~\bibnamefont {Delor}},\ }\href@noop {}
  {\bibfield  {journal} {\bibinfo  {journal} {J.\ Am.\ Chem.\ Soc.}\ }\textbf
  {\bibinfo {volume} {146}},\ \bibinfo {pages} {9544} (\bibinfo {year}
  {2024})}\BibitemShut {NoStop}%
\bibitem [{\citenamefont {Chervy}\ \emph {et~al.}(2016)\citenamefont {Chervy},
  \citenamefont {Xu}, \citenamefont {Duan}, \citenamefont {Wang}, \citenamefont
  {Mager}, \citenamefont {Frerejean}, \citenamefont {Munninghoff},
  \citenamefont {Tinnemans}, \citenamefont {Hutchison}, \citenamefont {Genet}
  \emph {et~al.}}]{chervy2016high}%
  \BibitemOpen
  \bibfield  {author} {\bibinfo {author} {\bibfnamefont {T.}~\bibnamefont
  {Chervy}}, \bibinfo {author} {\bibfnamefont {J.}~\bibnamefont {Xu}}, \bibinfo
  {author} {\bibfnamefont {Y.}~\bibnamefont {Duan}}, \bibinfo {author}
  {\bibfnamefont {C.}~\bibnamefont {Wang}}, \bibinfo {author} {\bibfnamefont
  {L.}~\bibnamefont {Mager}}, \bibinfo {author} {\bibfnamefont
  {M.}~\bibnamefont {Frerejean}}, \bibinfo {author} {\bibfnamefont {J.~A.}\
  \bibnamefont {Munninghoff}}, \bibinfo {author} {\bibfnamefont
  {P.}~\bibnamefont {Tinnemans}}, \bibinfo {author} {\bibfnamefont {J.~A.}\
  \bibnamefont {Hutchison}}, \bibinfo {author} {\bibfnamefont {C.}~\bibnamefont
  {Genet}}, \emph {et~al.},\ }\href@noop {} {\bibfield  {journal} {\bibinfo
  {journal} {Nano Lett.}\ }\textbf {\bibinfo {volume} {16}},\ \bibinfo {pages}
  {7352} (\bibinfo {year} {2016})}\BibitemShut {NoStop}%
\bibitem [{\citenamefont {Barachati}\ \emph {et~al.}(2018)\citenamefont
  {Barachati}, \citenamefont {Simon}, \citenamefont {Getmanenko}, \citenamefont
  {Barlow}, \citenamefont {Marder},\ and\ \citenamefont
  {K{\'e}na-Cohen}}]{barachati2018tunable}%
  \BibitemOpen
  \bibfield  {author} {\bibinfo {author} {\bibfnamefont {F.}~\bibnamefont
  {Barachati}}, \bibinfo {author} {\bibfnamefont {J.}~\bibnamefont {Simon}},
  \bibinfo {author} {\bibfnamefont {Y.~A.}\ \bibnamefont {Getmanenko}},
  \bibinfo {author} {\bibfnamefont {S.}~\bibnamefont {Barlow}}, \bibinfo
  {author} {\bibfnamefont {S.~R.}\ \bibnamefont {Marder}},\ and\ \bibinfo
  {author} {\bibfnamefont {S.}~\bibnamefont {K{\'e}na-Cohen}},\ }\href@noop {}
  {\bibfield  {journal} {\bibinfo  {journal} {ACS Photonics}\ }\textbf
  {\bibinfo {volume} {5}},\ \bibinfo {pages} {119} (\bibinfo {year}
  {2018})}\BibitemShut {NoStop}%
\bibitem [{\citenamefont {K{\'e}na-Cohen}\ and\ \citenamefont
  {Forrest}(2010)}]{kena2010room}%
  \BibitemOpen
  \bibfield  {author} {\bibinfo {author} {\bibfnamefont {S.}~\bibnamefont
  {K{\'e}na-Cohen}}\ and\ \bibinfo {author} {\bibfnamefont {S.}~\bibnamefont
  {Forrest}},\ }\href@noop {} {\bibfield  {journal} {\bibinfo  {journal} {Nat.\
  Photonics}\ }\textbf {\bibinfo {volume} {4}},\ \bibinfo {pages} {371}
  (\bibinfo {year} {2010})}\BibitemShut {NoStop}%
\bibitem [{\citenamefont {Daskalakis}\ \emph {et~al.}(2014)\citenamefont
  {Daskalakis}, \citenamefont {Maier}, \citenamefont {Murray},\ and\
  \citenamefont {K{\'e}na-Cohen}}]{daskalakis2014nonlinear}%
  \BibitemOpen
  \bibfield  {author} {\bibinfo {author} {\bibfnamefont {K.}~\bibnamefont
  {Daskalakis}}, \bibinfo {author} {\bibfnamefont {S.}~\bibnamefont {Maier}},
  \bibinfo {author} {\bibfnamefont {R.}~\bibnamefont {Murray}},\ and\ \bibinfo
  {author} {\bibfnamefont {S.}~\bibnamefont {K{\'e}na-Cohen}},\ }\href@noop {}
  {\bibfield  {journal} {\bibinfo  {journal} {Nat.\ Mater.}\ }\textbf {\bibinfo
  {volume} {13}},\ \bibinfo {pages} {271} (\bibinfo {year} {2014})}\BibitemShut
  {NoStop}%
\bibitem [{\citenamefont {Plumhof}\ \emph {et~al.}(2014)\citenamefont
  {Plumhof}, \citenamefont {St{\"o}ferle}, \citenamefont {Mai}, \citenamefont
  {Scherf},\ and\ \citenamefont {Mahrt}}]{plumhof2014room}%
  \BibitemOpen
  \bibfield  {author} {\bibinfo {author} {\bibfnamefont {J.~D.}\ \bibnamefont
  {Plumhof}}, \bibinfo {author} {\bibfnamefont {T.}~\bibnamefont
  {St{\"o}ferle}}, \bibinfo {author} {\bibfnamefont {L.}~\bibnamefont {Mai}},
  \bibinfo {author} {\bibfnamefont {U.}~\bibnamefont {Scherf}},\ and\ \bibinfo
  {author} {\bibfnamefont {R.~F.}\ \bibnamefont {Mahrt}},\ }\href@noop {}
  {\bibfield  {journal} {\bibinfo  {journal} {Nat.\ Mater.}\ }\textbf {\bibinfo
  {volume} {13}},\ \bibinfo {pages} {247} (\bibinfo {year} {2014})}\BibitemShut
  {NoStop}%
\bibitem [{\citenamefont {Dovzhenko}\ \emph {et~al.}(2018)\citenamefont
  {Dovzhenko}, \citenamefont {Ryabchuk}, \citenamefont {Rakovich},\ and\
  \citenamefont {Nabiev}}]{dovzhenko2018light}%
  \BibitemOpen
  \bibfield  {author} {\bibinfo {author} {\bibfnamefont {D.}~\bibnamefont
  {Dovzhenko}}, \bibinfo {author} {\bibfnamefont {S.}~\bibnamefont {Ryabchuk}},
  \bibinfo {author} {\bibfnamefont {Y.~P.}\ \bibnamefont {Rakovich}},\ and\
  \bibinfo {author} {\bibfnamefont {I.}~\bibnamefont {Nabiev}},\ }\href@noop {}
  {\bibfield  {journal} {\bibinfo  {journal} {Nanoscale}\ }\textbf {\bibinfo
  {volume} {10}},\ \bibinfo {pages} {3589} (\bibinfo {year}
  {2018})}\BibitemShut {NoStop}%
\bibitem [{\citenamefont {Bhuyan}\ \emph {et~al.}(2023)\citenamefont {Bhuyan},
  \citenamefont {Mony}, \citenamefont {Kotov}, \citenamefont {Castellanos},
  \citenamefont {G{\'o}mez~Rivas}, \citenamefont {Shegai},\ and\ \citenamefont
  {B{\"o}rjesson}}]{bhuyan2023rise}%
  \BibitemOpen
  \bibfield  {author} {\bibinfo {author} {\bibfnamefont {R.}~\bibnamefont
  {Bhuyan}}, \bibinfo {author} {\bibfnamefont {J.}~\bibnamefont {Mony}},
  \bibinfo {author} {\bibfnamefont {O.}~\bibnamefont {Kotov}}, \bibinfo
  {author} {\bibfnamefont {G.~W.}\ \bibnamefont {Castellanos}}, \bibinfo
  {author} {\bibfnamefont {J.}~\bibnamefont {G{\'o}mez~Rivas}}, \bibinfo
  {author} {\bibfnamefont {T.~O.}\ \bibnamefont {Shegai}},\ and\ \bibinfo
  {author} {\bibfnamefont {K.}~\bibnamefont {B{\"o}rjesson}},\ }\href@noop {}
  {\bibfield  {journal} {\bibinfo  {journal} {Chem.\ Rev.}\ }\textbf {\bibinfo
  {volume} {123}},\ \bibinfo {pages} {10877} (\bibinfo {year}
  {2023})}\BibitemShut {NoStop}%
\bibitem [{\citenamefont {Simpkins}\ \emph {et~al.}(2015)\citenamefont
  {Simpkins}, \citenamefont {Fears}, \citenamefont {Dressick}, \citenamefont
  {Spann}, \citenamefont {Dunkelberger},\ and\ \citenamefont
  {Owrutsky}}]{simpkins2015spanning}%
  \BibitemOpen
  \bibfield  {author} {\bibinfo {author} {\bibfnamefont {B.}~\bibnamefont
  {Simpkins}}, \bibinfo {author} {\bibfnamefont {K.~P.}\ \bibnamefont {Fears}},
  \bibinfo {author} {\bibfnamefont {W.~J.}\ \bibnamefont {Dressick}}, \bibinfo
  {author} {\bibfnamefont {B.~T.}\ \bibnamefont {Spann}}, \bibinfo {author}
  {\bibfnamefont {A.~D.}\ \bibnamefont {Dunkelberger}},\ and\ \bibinfo {author}
  {\bibfnamefont {J.~C.}\ \bibnamefont {Owrutsky}},\ }\href@noop {} {\bibfield
  {journal} {\bibinfo  {journal} {ACS Photonics}\ }\textbf {\bibinfo {volume}
  {2}},\ \bibinfo {pages} {1460} (\bibinfo {year} {2015})}\BibitemShut
  {NoStop}%
\bibitem [{\citenamefont {Wright}\ \emph {et~al.}(2023)\citenamefont {Wright},
  \citenamefont {Nelson},\ and\ \citenamefont
  {Weichman}}]{wright2023rovibrational}%
  \BibitemOpen
  \bibfield  {author} {\bibinfo {author} {\bibfnamefont {A.~D.}\ \bibnamefont
  {Wright}}, \bibinfo {author} {\bibfnamefont {J.~C.}\ \bibnamefont {Nelson}},\
  and\ \bibinfo {author} {\bibfnamefont {M.~L.}\ \bibnamefont {Weichman}},\
  }\href@noop {} {\bibfield  {journal} {\bibinfo  {journal} {J.\ Am.\ Chem.\
  Soc.}\ }\textbf {\bibinfo {volume} {145}},\ \bibinfo {pages} {5982} (\bibinfo
  {year} {2023})}\BibitemShut {NoStop}%
\bibitem [{\citenamefont {Mandal}\ \emph {et~al.}(2023)\citenamefont {Mandal},
  \citenamefont {Taylor}, \citenamefont {Weight}, \citenamefont {Koessler},
  \citenamefont {Li},\ and\ \citenamefont {Huo}}]{mandal2023theoretical}%
  \BibitemOpen
  \bibfield  {author} {\bibinfo {author} {\bibfnamefont {A.}~\bibnamefont
  {Mandal}}, \bibinfo {author} {\bibfnamefont {M.~A.}\ \bibnamefont {Taylor}},
  \bibinfo {author} {\bibfnamefont {B.~M.}\ \bibnamefont {Weight}}, \bibinfo
  {author} {\bibfnamefont {E.~R.}\ \bibnamefont {Koessler}}, \bibinfo {author}
  {\bibfnamefont {X.}~\bibnamefont {Li}},\ and\ \bibinfo {author}
  {\bibfnamefont {P.}~\bibnamefont {Huo}},\ }\href@noop {} {\bibfield
  {journal} {\bibinfo  {journal} {Chem.\ Rev.}\ }\textbf {\bibinfo {volume}
  {123}},\ \bibinfo {pages} {9786} (\bibinfo {year} {2023})}\BibitemShut
  {NoStop}%
\bibitem [{\citenamefont {Agranovich}\ \emph {et~al.}(2003)\citenamefont
  {Agranovich}, \citenamefont {Litinskaia},\ and\ \citenamefont
  {Lidzey}}]{agranovich2003cavity}%
  \BibitemOpen
  \bibfield  {author} {\bibinfo {author} {\bibfnamefont {V.~M.}\ \bibnamefont
  {Agranovich}}, \bibinfo {author} {\bibfnamefont {M.}~\bibnamefont
  {Litinskaia}},\ and\ \bibinfo {author} {\bibfnamefont {D.~G.}\ \bibnamefont
  {Lidzey}},\ }\href@noop {} {\bibfield  {journal} {\bibinfo  {journal} {Phys.\
  Rev.\ B}\ }\textbf {\bibinfo {volume} {67}},\ \bibinfo {pages} {085311}
  (\bibinfo {year} {2003})}\BibitemShut {NoStop}%
\bibitem [{\citenamefont {Litinskaya}\ \emph {et~al.}(2004)\citenamefont
  {Litinskaya}, \citenamefont {Reineker},\ and\ \citenamefont
  {Agranovich}}]{litinskaya2004fast}%
  \BibitemOpen
  \bibfield  {author} {\bibinfo {author} {\bibfnamefont {M.}~\bibnamefont
  {Litinskaya}}, \bibinfo {author} {\bibfnamefont {P.}~\bibnamefont
  {Reineker}},\ and\ \bibinfo {author} {\bibfnamefont {V.~M.}\ \bibnamefont
  {Agranovich}},\ }\href@noop {} {\bibfield  {journal} {\bibinfo  {journal}
  {J.\ Lumin.}\ }\textbf {\bibinfo {volume} {110}},\ \bibinfo {pages} {364}
  (\bibinfo {year} {2004})}\BibitemShut {NoStop}%
\bibitem [{\citenamefont {Virgili}\ \emph {et~al.}(2011)\citenamefont
  {Virgili}, \citenamefont {Coles}, \citenamefont {Adawi}, \citenamefont
  {Clark}, \citenamefont {Michetti}, \citenamefont {Rajendran}, \citenamefont
  {Brida}, \citenamefont {Polli}, \citenamefont {Cerullo},\ and\ \citenamefont
  {Lidzey}}]{virgili2011ultrafast}%
  \BibitemOpen
  \bibfield  {author} {\bibinfo {author} {\bibfnamefont {T.}~\bibnamefont
  {Virgili}}, \bibinfo {author} {\bibfnamefont {D.}~\bibnamefont {Coles}},
  \bibinfo {author} {\bibfnamefont {A.}~\bibnamefont {Adawi}}, \bibinfo
  {author} {\bibfnamefont {C.}~\bibnamefont {Clark}}, \bibinfo {author}
  {\bibfnamefont {P.}~\bibnamefont {Michetti}}, \bibinfo {author}
  {\bibfnamefont {S.}~\bibnamefont {Rajendran}}, \bibinfo {author}
  {\bibfnamefont {D.}~\bibnamefont {Brida}}, \bibinfo {author} {\bibfnamefont
  {D.}~\bibnamefont {Polli}}, \bibinfo {author} {\bibfnamefont
  {G.}~\bibnamefont {Cerullo}},\ and\ \bibinfo {author} {\bibfnamefont
  {D.}~\bibnamefont {Lidzey}},\ }\href@noop {} {\bibfield  {journal} {\bibinfo
  {journal} {Phys.\ Rev.\ B}\ }\textbf {\bibinfo {volume} {83}},\ \bibinfo
  {pages} {245309} (\bibinfo {year} {2011})}\BibitemShut {NoStop}%
\bibitem [{\citenamefont {Canaguier-Durand}\ \emph {et~al.}(2013)\citenamefont
  {Canaguier-Durand}, \citenamefont {Devaux}, \citenamefont {George},
  \citenamefont {Pang}, \citenamefont {Hutchison}, \citenamefont {Schwartz},
  \citenamefont {Genet}, \citenamefont {Wilhelms}, \citenamefont {Lehn},\ and\
  \citenamefont {Ebbesen}}]{canaguier2013thermodynamics}%
  \BibitemOpen
  \bibfield  {author} {\bibinfo {author} {\bibfnamefont {A.}~\bibnamefont
  {Canaguier-Durand}}, \bibinfo {author} {\bibfnamefont {E.}~\bibnamefont
  {Devaux}}, \bibinfo {author} {\bibfnamefont {J.}~\bibnamefont {George}},
  \bibinfo {author} {\bibfnamefont {Y.}~\bibnamefont {Pang}}, \bibinfo {author}
  {\bibfnamefont {J.~A.}\ \bibnamefont {Hutchison}}, \bibinfo {author}
  {\bibfnamefont {T.}~\bibnamefont {Schwartz}}, \bibinfo {author}
  {\bibfnamefont {C.}~\bibnamefont {Genet}}, \bibinfo {author} {\bibfnamefont
  {N.}~\bibnamefont {Wilhelms}}, \bibinfo {author} {\bibfnamefont {J.-M.}\
  \bibnamefont {Lehn}},\ and\ \bibinfo {author} {\bibfnamefont {T.~W.}\
  \bibnamefont {Ebbesen}},\ }\href@noop {} {\bibfield  {journal} {\bibinfo
  {journal} {Angew.\ Chem.\ Int.\ Ed.}\ }\textbf {\bibinfo {volume} {52}},\
  \bibinfo {pages} {10533} (\bibinfo {year} {2013})}\BibitemShut {NoStop}%
\bibitem [{\citenamefont {Wang}\ \emph {et~al.}(2014)\citenamefont {Wang},
  \citenamefont {Chervy}, \citenamefont {George}, \citenamefont {Hutchison},
  \citenamefont {Genet},\ and\ \citenamefont {Ebbesen}}]{wang2014quantum}%
  \BibitemOpen
  \bibfield  {author} {\bibinfo {author} {\bibfnamefont {S.}~\bibnamefont
  {Wang}}, \bibinfo {author} {\bibfnamefont {T.}~\bibnamefont {Chervy}},
  \bibinfo {author} {\bibfnamefont {J.}~\bibnamefont {George}}, \bibinfo
  {author} {\bibfnamefont {J.~A.}\ \bibnamefont {Hutchison}}, \bibinfo {author}
  {\bibfnamefont {C.}~\bibnamefont {Genet}},\ and\ \bibinfo {author}
  {\bibfnamefont {T.~W.}\ \bibnamefont {Ebbesen}},\ }\href@noop {} {\bibfield
  {journal} {\bibinfo  {journal} {J.\ Phys.\ Chem.\ Lett.}\ }\textbf {\bibinfo
  {volume} {5}},\ \bibinfo {pages} {1433} (\bibinfo {year} {2014})}\BibitemShut
  {NoStop}%
\bibitem [{\citenamefont {Shalabney}\ \emph {et~al.}(2015)\citenamefont
  {Shalabney}, \citenamefont {George}, \citenamefont {Hiura}, \citenamefont
  {Hutchison}, \citenamefont {Genet}, \citenamefont {Hellwig},\ and\
  \citenamefont {Ebbesen}}]{shalabney2015enhanced}%
  \BibitemOpen
  \bibfield  {author} {\bibinfo {author} {\bibfnamefont {A.}~\bibnamefont
  {Shalabney}}, \bibinfo {author} {\bibfnamefont {J.}~\bibnamefont {George}},
  \bibinfo {author} {\bibfnamefont {H.}~\bibnamefont {Hiura}}, \bibinfo
  {author} {\bibfnamefont {J.~A.}\ \bibnamefont {Hutchison}}, \bibinfo {author}
  {\bibfnamefont {C.}~\bibnamefont {Genet}}, \bibinfo {author} {\bibfnamefont
  {P.}~\bibnamefont {Hellwig}},\ and\ \bibinfo {author} {\bibfnamefont {T.~W.}\
  \bibnamefont {Ebbesen}},\ }\href@noop {} {\bibfield  {journal} {\bibinfo
  {journal} {Angew.\ Chem.\ Int.\ Ed.}\ }\textbf {\bibinfo {volume} {54}},\
  \bibinfo {pages} {7971} (\bibinfo {year} {2015})}\BibitemShut {NoStop}%
\bibitem [{\citenamefont {Ahn}\ \emph {et~al.}(2018)\citenamefont {Ahn},
  \citenamefont {Vurgaftman}, \citenamefont {Dunkelberger}, \citenamefont
  {Owrutsky},\ and\ \citenamefont {Simpkins}}]{ahn2018vibrational}%
  \BibitemOpen
  \bibfield  {author} {\bibinfo {author} {\bibfnamefont {W.}~\bibnamefont
  {Ahn}}, \bibinfo {author} {\bibfnamefont {I.}~\bibnamefont {Vurgaftman}},
  \bibinfo {author} {\bibfnamefont {A.~D.}\ \bibnamefont {Dunkelberger}},
  \bibinfo {author} {\bibfnamefont {J.~C.}\ \bibnamefont {Owrutsky}},\ and\
  \bibinfo {author} {\bibfnamefont {B.~S.}\ \bibnamefont {Simpkins}},\
  }\href@noop {} {\bibfield  {journal} {\bibinfo  {journal} {ACS Photonics}\
  }\textbf {\bibinfo {volume} {5}},\ \bibinfo {pages} {158} (\bibinfo {year}
  {2018})}\BibitemShut {NoStop}%
\bibitem [{\citenamefont {DelPo}\ \emph {et~al.}(2021)\citenamefont {DelPo},
  \citenamefont {Khan}, \citenamefont {Park}, \citenamefont {Kudisch},
  \citenamefont {Rand},\ and\ \citenamefont {Scholes}}]{delpo2021polariton}%
  \BibitemOpen
  \bibfield  {author} {\bibinfo {author} {\bibfnamefont {C.~A.}\ \bibnamefont
  {DelPo}}, \bibinfo {author} {\bibfnamefont {S.-U.-Z.}\ \bibnamefont {Khan}},
  \bibinfo {author} {\bibfnamefont {K.~H.}\ \bibnamefont {Park}}, \bibinfo
  {author} {\bibfnamefont {B.}~\bibnamefont {Kudisch}}, \bibinfo {author}
  {\bibfnamefont {B.~P.}\ \bibnamefont {Rand}},\ and\ \bibinfo {author}
  {\bibfnamefont {G.~D.}\ \bibnamefont {Scholes}},\ }\href@noop {} {\bibfield
  {journal} {\bibinfo  {journal} {J.\ Phys.\ Chem.\ Lett.}\ }\textbf {\bibinfo
  {volume} {12}},\ \bibinfo {pages} {9774} (\bibinfo {year}
  {2021})}\BibitemShut {NoStop}%
\bibitem [{\citenamefont {McKillop}\ and\ \citenamefont
  {Weichman}(2025)}]{mckillop2025cavity}%
  \BibitemOpen
  \bibfield  {author} {\bibinfo {author} {\bibfnamefont {A.~M.}\ \bibnamefont
  {McKillop}}\ and\ \bibinfo {author} {\bibfnamefont {M.~L.}\ \bibnamefont
  {Weichman}},\ }\href@noop {} {\bibfield  {journal} {\bibinfo  {journal}
  {Chem.\ Phys.\ Rev.}\ }\textbf {\bibinfo {volume} {6}},\ \bibinfo {pages}
  {031308} (\bibinfo {year} {2025})}\BibitemShut {NoStop}%
\bibitem [{\citenamefont {Michetti}\ and\ \citenamefont
  {La~Rocca}(2005)}]{michetti2005polariton}%
  \BibitemOpen
  \bibfield  {author} {\bibinfo {author} {\bibfnamefont {P.}~\bibnamefont
  {Michetti}}\ and\ \bibinfo {author} {\bibfnamefont {G.~C.}\ \bibnamefont
  {La~Rocca}},\ }\href@noop {} {\bibfield  {journal} {\bibinfo  {journal}
  {Phys.\ Rev.\ B}\ }\textbf {\bibinfo {volume} {71}},\ \bibinfo {pages}
  {115320} (\bibinfo {year} {2005})}\BibitemShut {NoStop}%
\bibitem [{\citenamefont {Stemo}\ \emph {et~al.}(2022)\citenamefont {Stemo},
  \citenamefont {Yamada}, \citenamefont {Katsuki},\ and\ \citenamefont
  {Yanagi}}]{stemo2022influence}%
  \BibitemOpen
  \bibfield  {author} {\bibinfo {author} {\bibfnamefont {G.}~\bibnamefont
  {Stemo}}, \bibinfo {author} {\bibfnamefont {H.}~\bibnamefont {Yamada}},
  \bibinfo {author} {\bibfnamefont {H.}~\bibnamefont {Katsuki}},\ and\ \bibinfo
  {author} {\bibfnamefont {H.}~\bibnamefont {Yanagi}},\ }\href@noop {}
  {\bibfield  {journal} {\bibinfo  {journal} {J.\ Phys.\ Chem.\ B}\ }\textbf
  {\bibinfo {volume} {126}},\ \bibinfo {pages} {9399} (\bibinfo {year}
  {2022})}\BibitemShut {NoStop}%
\bibitem [{\citenamefont {Liu}\ and\ \citenamefont
  {Holmes}(2025)}]{liu2025continuously}%
  \BibitemOpen
  \bibfield  {author} {\bibinfo {author} {\bibfnamefont {Y.}~\bibnamefont
  {Liu}}\ and\ \bibinfo {author} {\bibfnamefont {R.~J.}\ \bibnamefont
  {Holmes}},\ }\href@noop {} {\bibfield  {journal} {\bibinfo  {journal} {Adv.\
  Opt.\ Mater.}\ }\textbf {\bibinfo {volume} {13}},\ \bibinfo {pages} {2403135}
  (\bibinfo {year} {2025})}\BibitemShut {NoStop}%
\bibitem [{\citenamefont {Schwartz}\ \emph {et~al.}(2013)\citenamefont
  {Schwartz}, \citenamefont {Hutchison}, \citenamefont {L{\'e}onard},
  \citenamefont {Genet}, \citenamefont {Haacke},\ and\ \citenamefont
  {Ebbesen}}]{schwartz2013polariton}%
  \BibitemOpen
  \bibfield  {author} {\bibinfo {author} {\bibfnamefont {T.}~\bibnamefont
  {Schwartz}}, \bibinfo {author} {\bibfnamefont {J.~A.}\ \bibnamefont
  {Hutchison}}, \bibinfo {author} {\bibfnamefont {J.}~\bibnamefont
  {L{\'e}onard}}, \bibinfo {author} {\bibfnamefont {C.}~\bibnamefont {Genet}},
  \bibinfo {author} {\bibfnamefont {S.}~\bibnamefont {Haacke}},\ and\ \bibinfo
  {author} {\bibfnamefont {T.~W.}\ \bibnamefont {Ebbesen}},\ }\href@noop {}
  {\bibfield  {journal} {\bibinfo  {journal} {ChemPhysChem}\ }\textbf {\bibinfo
  {volume} {14}},\ \bibinfo {pages} {125} (\bibinfo {year} {2013})}\BibitemShut
  {NoStop}%
\bibitem [{\citenamefont {Xiang}\ \emph {et~al.}(2018)\citenamefont {Xiang},
  \citenamefont {Ribeiro}, \citenamefont {Dunkelberger}, \citenamefont {Wang},
  \citenamefont {Li}, \citenamefont {Simpkins}, \citenamefont {Owrutsky},
  \citenamefont {Yuen-Zhou},\ and\ \citenamefont {Xiong}}]{xiang2018two}%
  \BibitemOpen
  \bibfield  {author} {\bibinfo {author} {\bibfnamefont {B.}~\bibnamefont
  {Xiang}}, \bibinfo {author} {\bibfnamefont {R.~F.}\ \bibnamefont {Ribeiro}},
  \bibinfo {author} {\bibfnamefont {A.~D.}\ \bibnamefont {Dunkelberger}},
  \bibinfo {author} {\bibfnamefont {J.}~\bibnamefont {Wang}}, \bibinfo {author}
  {\bibfnamefont {Y.}~\bibnamefont {Li}}, \bibinfo {author} {\bibfnamefont
  {B.~S.}\ \bibnamefont {Simpkins}}, \bibinfo {author} {\bibfnamefont {J.~C.}\
  \bibnamefont {Owrutsky}}, \bibinfo {author} {\bibfnamefont {J.}~\bibnamefont
  {Yuen-Zhou}},\ and\ \bibinfo {author} {\bibfnamefont {W.}~\bibnamefont
  {Xiong}},\ }\href@noop {} {\bibfield  {journal} {\bibinfo  {journal} {Proc.\
  Natl.\ Acad.\ Sci.}\ }\textbf {\bibinfo {volume} {115}},\ \bibinfo {pages}
  {4845} (\bibinfo {year} {2018})}\BibitemShut {NoStop}%
\bibitem [{\citenamefont {Renken}\ \emph {et~al.}(2021)\citenamefont {Renken},
  \citenamefont {Pandya}, \citenamefont {Georgiou}, \citenamefont
  {Jayaprakash}, \citenamefont {Gai}, \citenamefont {Shen}, \citenamefont
  {Lidzey}, \citenamefont {Rao},\ and\ \citenamefont
  {Musser}}]{renken2021untargeted}%
  \BibitemOpen
  \bibfield  {author} {\bibinfo {author} {\bibfnamefont {S.}~\bibnamefont
  {Renken}}, \bibinfo {author} {\bibfnamefont {R.}~\bibnamefont {Pandya}},
  \bibinfo {author} {\bibfnamefont {K.}~\bibnamefont {Georgiou}}, \bibinfo
  {author} {\bibfnamefont {R.}~\bibnamefont {Jayaprakash}}, \bibinfo {author}
  {\bibfnamefont {L.}~\bibnamefont {Gai}}, \bibinfo {author} {\bibfnamefont
  {Z.}~\bibnamefont {Shen}}, \bibinfo {author} {\bibfnamefont {D.~G.}\
  \bibnamefont {Lidzey}}, \bibinfo {author} {\bibfnamefont {A.}~\bibnamefont
  {Rao}},\ and\ \bibinfo {author} {\bibfnamefont {A.~J.}\ \bibnamefont
  {Musser}},\ }\href@noop {} {\bibfield  {journal} {\bibinfo  {journal} {J.\
  Chem.\ Phys.}\ }\textbf {\bibinfo {volume} {155}},\ \bibinfo {pages} {154701}
  (\bibinfo {year} {2021})}\BibitemShut {NoStop}%
\bibitem [{\citenamefont {Pyles}\ \emph {et~al.}(2024)\citenamefont {Pyles},
  \citenamefont {Simpkins}, \citenamefont {Vurgaftman}, \citenamefont
  {Owrutsky},\ and\ \citenamefont {Dunkelberger}}]{pyles2024revisiting}%
  \BibitemOpen
  \bibfield  {author} {\bibinfo {author} {\bibfnamefont {C.~G.}\ \bibnamefont
  {Pyles}}, \bibinfo {author} {\bibfnamefont {B.~S.}\ \bibnamefont {Simpkins}},
  \bibinfo {author} {\bibfnamefont {I.}~\bibnamefont {Vurgaftman}}, \bibinfo
  {author} {\bibfnamefont {J.~C.}\ \bibnamefont {Owrutsky}},\ and\ \bibinfo
  {author} {\bibfnamefont {A.~D.}\ \bibnamefont {Dunkelberger}},\ }\href@noop
  {} {\bibfield  {journal} {\bibinfo  {journal} {J.\ Chem.\ Phys.}\ }\textbf
  {\bibinfo {volume} {161}},\ \bibinfo {pages} {234202} (\bibinfo {year}
  {2024})}\BibitemShut {NoStop}%
\bibitem [{\citenamefont {Schwennicke}\ \emph {et~al.}(2025)\citenamefont
  {Schwennicke}, \citenamefont {Koner}, \citenamefont {P{\'e}rez-S{\'a}nchez},
  \citenamefont {Xiong}, \citenamefont {Giebink}, \citenamefont {Weichman},\
  and\ \citenamefont {Yuen-Zhou}}]{schwennicke2025molecular}%
  \BibitemOpen
  \bibfield  {author} {\bibinfo {author} {\bibfnamefont {K.}~\bibnamefont
  {Schwennicke}}, \bibinfo {author} {\bibfnamefont {A.}~\bibnamefont {Koner}},
  \bibinfo {author} {\bibfnamefont {J.~B.}\ \bibnamefont
  {P{\'e}rez-S{\'a}nchez}}, \bibinfo {author} {\bibfnamefont {W.}~\bibnamefont
  {Xiong}}, \bibinfo {author} {\bibfnamefont {N.~C.}\ \bibnamefont {Giebink}},
  \bibinfo {author} {\bibfnamefont {M.~L.}\ \bibnamefont {Weichman}},\ and\
  \bibinfo {author} {\bibfnamefont {J.}~\bibnamefont {Yuen-Zhou}},\ }\href@noop
  {} {\bibfield  {journal} {\bibinfo  {journal} {Chem.\ Soc.\ Rev.}\ }\textbf
  {\bibinfo {volume} {54}},\ \bibinfo {pages} {6482} (\bibinfo {year}
  {2025})}\BibitemShut {NoStop}%
\bibitem [{\citenamefont {McKillop}\ \emph {et~al.}(2026)\citenamefont
  {McKillop}, \citenamefont {Chen}, \citenamefont {Fidler},\ and\ \citenamefont
  {Weichman}}]{mckillop2026direct}%
  \BibitemOpen
  \bibfield  {author} {\bibinfo {author} {\bibfnamefont {A.~M.}\ \bibnamefont
  {McKillop}}, \bibinfo {author} {\bibfnamefont {L.}~\bibnamefont {Chen}},
  \bibinfo {author} {\bibfnamefont {A.~P.}\ \bibnamefont {Fidler}},\ and\
  \bibinfo {author} {\bibfnamefont {M.~L.}\ \bibnamefont {Weichman}},\
  }\href@noop {} {\bibfield  {journal} {\bibinfo  {journal} {J.\ Am.\ Chem.\
  Soc.}\ }\textbf {\bibinfo {volume} {148}},\ \bibinfo {pages} {9739} (\bibinfo
  {year} {2026})}\BibitemShut {NoStop}%
\bibitem [{\citenamefont {Song}\ \emph {et~al.}(2004)\citenamefont {Song},
  \citenamefont {He}, \citenamefont {Nurmikko}, \citenamefont {Tischler},\ and\
  \citenamefont {Bulovic}}]{song2004exciton}%
  \BibitemOpen
  \bibfield  {author} {\bibinfo {author} {\bibfnamefont {J.-H.}\ \bibnamefont
  {Song}}, \bibinfo {author} {\bibfnamefont {Y.}~\bibnamefont {He}}, \bibinfo
  {author} {\bibfnamefont {A.}~\bibnamefont {Nurmikko}}, \bibinfo {author}
  {\bibfnamefont {J.}~\bibnamefont {Tischler}},\ and\ \bibinfo {author}
  {\bibfnamefont {V.}~\bibnamefont {Bulovic}},\ }\href@noop {} {\bibfield
  {journal} {\bibinfo  {journal} {Phys.\ Rev.\ B}\ }\textbf {\bibinfo {volume}
  {69}},\ \bibinfo {pages} {235330} (\bibinfo {year} {2004})}\BibitemShut
  {NoStop}%
\bibitem [{\citenamefont {Avramenko}\ and\ \citenamefont
  {Rury}(2020)}]{avramenko2020quantum}%
  \BibitemOpen
  \bibfield  {author} {\bibinfo {author} {\bibfnamefont {A.~G.}\ \bibnamefont
  {Avramenko}}\ and\ \bibinfo {author} {\bibfnamefont {A.~S.}\ \bibnamefont
  {Rury}},\ }\href@noop {} {\bibfield  {journal} {\bibinfo  {journal} {J.\
  Phys.\ Chem.\ Lett.}\ }\textbf {\bibinfo {volume} {11}},\ \bibinfo {pages}
  {1013} (\bibinfo {year} {2020})}\BibitemShut {NoStop}%
\bibitem [{\citenamefont {Wu}\ \emph {et~al.}(2022)\citenamefont {Wu},
  \citenamefont {Finkelstein-Shapiro}, \citenamefont {Wang}, \citenamefont
  {Rosenkampff}, \citenamefont {Yartsev}, \citenamefont {Pascher},
  \citenamefont {Nguyen-Phan}, \citenamefont {Cogdell}, \citenamefont
  {B{\"o}rjesson},\ and\ \citenamefont {Pullerits}}]{wu2022optical}%
  \BibitemOpen
  \bibfield  {author} {\bibinfo {author} {\bibfnamefont {F.}~\bibnamefont
  {Wu}}, \bibinfo {author} {\bibfnamefont {D.}~\bibnamefont
  {Finkelstein-Shapiro}}, \bibinfo {author} {\bibfnamefont {M.}~\bibnamefont
  {Wang}}, \bibinfo {author} {\bibfnamefont {I.}~\bibnamefont {Rosenkampff}},
  \bibinfo {author} {\bibfnamefont {A.}~\bibnamefont {Yartsev}}, \bibinfo
  {author} {\bibfnamefont {T.}~\bibnamefont {Pascher}}, \bibinfo {author}
  {\bibfnamefont {T.~C.}\ \bibnamefont {Nguyen-Phan}}, \bibinfo {author}
  {\bibfnamefont {R.}~\bibnamefont {Cogdell}}, \bibinfo {author} {\bibfnamefont
  {K.}~\bibnamefont {B{\"o}rjesson}},\ and\ \bibinfo {author} {\bibfnamefont
  {T.}~\bibnamefont {Pullerits}},\ }\href@noop {} {\bibfield  {journal}
  {\bibinfo  {journal} {Nat.\ Commun.}\ }\textbf {\bibinfo {volume} {13}},\
  \bibinfo {pages} {6864} (\bibinfo {year} {2022})}\BibitemShut {NoStop}%
\bibitem [{\citenamefont {Kushida}\ \emph {et~al.}(2025)\citenamefont
  {Kushida}, \citenamefont {Wang}, \citenamefont {Seidel}, \citenamefont
  {Yamamoto}, \citenamefont {Genet},\ and\ \citenamefont
  {Ebbesen}}]{kushida2025fluidic}%
  \BibitemOpen
  \bibfield  {author} {\bibinfo {author} {\bibfnamefont {S.}~\bibnamefont
  {Kushida}}, \bibinfo {author} {\bibfnamefont {K.}~\bibnamefont {Wang}},
  \bibinfo {author} {\bibfnamefont {M.}~\bibnamefont {Seidel}}, \bibinfo
  {author} {\bibfnamefont {Y.}~\bibnamefont {Yamamoto}}, \bibinfo {author}
  {\bibfnamefont {C.}~\bibnamefont {Genet}},\ and\ \bibinfo {author}
  {\bibfnamefont {T.~W.}\ \bibnamefont {Ebbesen}},\ }\href@noop {} {\bibfield
  {journal} {\bibinfo  {journal} {J.\ Phys.\ Chem.\ Lett.}\ }\textbf {\bibinfo
  {volume} {16}},\ \bibinfo {pages} {8570} (\bibinfo {year}
  {2025})}\BibitemShut {NoStop}%
\bibitem [{\citenamefont {Rashidi}\ \emph {et~al.}(2025)\citenamefont
  {Rashidi}, \citenamefont {Michail}, \citenamefont {Salcido-Santacruz},
  \citenamefont {Paudel}, \citenamefont {Menon},\ and\ \citenamefont
  {Sfeir}}]{rashidi2025efficient}%
  \BibitemOpen
  \bibfield  {author} {\bibinfo {author} {\bibfnamefont {K.}~\bibnamefont
  {Rashidi}}, \bibinfo {author} {\bibfnamefont {E.}~\bibnamefont {Michail}},
  \bibinfo {author} {\bibfnamefont {B.}~\bibnamefont {Salcido-Santacruz}},
  \bibinfo {author} {\bibfnamefont {Y.}~\bibnamefont {Paudel}}, \bibinfo
  {author} {\bibfnamefont {V.~M.}\ \bibnamefont {Menon}},\ and\ \bibinfo
  {author} {\bibfnamefont {M.~Y.}\ \bibnamefont {Sfeir}},\ }\href@noop {}
  {\bibfield  {journal} {\bibinfo  {journal} {Nat.\ Nanotechnol.}\ }\textbf
  {\bibinfo {volume} {20}},\ \bibinfo {pages} {1618} (\bibinfo {year}
  {2025})}\BibitemShut {NoStop}%
\bibitem [{\citenamefont {Liu}\ \emph {et~al.}(2021)\citenamefont {Liu},
  \citenamefont {Menon},\ and\ \citenamefont {Sfeir}}]{liu2021ultrafast}%
  \BibitemOpen
  \bibfield  {author} {\bibinfo {author} {\bibfnamefont {B.}~\bibnamefont
  {Liu}}, \bibinfo {author} {\bibfnamefont {V.~M.}\ \bibnamefont {Menon}},\
  and\ \bibinfo {author} {\bibfnamefont {M.~Y.}\ \bibnamefont {Sfeir}},\
  }\href@noop {} {\bibfield  {journal} {\bibinfo  {journal} {APL Photonics}\
  }\textbf {\bibinfo {volume} {6}},\ \bibinfo {pages} {016103} (\bibinfo {year}
  {2021})}\BibitemShut {NoStop}%
\bibitem [{\citenamefont {Fidler}\ \emph {et~al.}(2023)\citenamefont {Fidler},
  \citenamefont {Chen}, \citenamefont {McKillop},\ and\ \citenamefont
  {Weichman}}]{fidler2023ultrafast}%
  \BibitemOpen
  \bibfield  {author} {\bibinfo {author} {\bibfnamefont {A.~P.}\ \bibnamefont
  {Fidler}}, \bibinfo {author} {\bibfnamefont {L.}~\bibnamefont {Chen}},
  \bibinfo {author} {\bibfnamefont {A.~M.}\ \bibnamefont {McKillop}},\ and\
  \bibinfo {author} {\bibfnamefont {M.~L.}\ \bibnamefont {Weichman}},\
  }\href@noop {} {\bibfield  {journal} {\bibinfo  {journal} {J.\ Chem.\ Phys.}\
  }\textbf {\bibinfo {volume} {159}},\ \bibinfo {pages} {164302} (\bibinfo
  {year} {2023})}\BibitemShut {NoStop}%
\bibitem [{\citenamefont {Chen}\ \emph {et~al.}(2024)\citenamefont {Chen},
  \citenamefont {Fidler}, \citenamefont {McKillop},\ and\ \citenamefont
  {Weichman}}]{chen2024exploring}%
  \BibitemOpen
  \bibfield  {author} {\bibinfo {author} {\bibfnamefont {L.}~\bibnamefont
  {Chen}}, \bibinfo {author} {\bibfnamefont {A.~P.}\ \bibnamefont {Fidler}},
  \bibinfo {author} {\bibfnamefont {A.~M.}\ \bibnamefont {McKillop}},\ and\
  \bibinfo {author} {\bibfnamefont {M.~L.}\ \bibnamefont {Weichman}},\
  }\href@noop {} {\bibfield  {journal} {\bibinfo  {journal} {Nanophotonics}\
  }\textbf {\bibinfo {volume} {13}},\ \bibinfo {pages} {2591} (\bibinfo {year}
  {2024})}\BibitemShut {NoStop}%
\bibitem [{\citenamefont {Avramenko}\ and\ \citenamefont
  {Rury}(2021)}]{avramenko2021local}%
  \BibitemOpen
  \bibfield  {author} {\bibinfo {author} {\bibfnamefont {A.~G.}\ \bibnamefont
  {Avramenko}}\ and\ \bibinfo {author} {\bibfnamefont {A.~S.}\ \bibnamefont
  {Rury}},\ }\href@noop {} {\bibfield  {journal} {\bibinfo  {journal} {J.\
  Chem.\ Phys.}\ }\textbf {\bibinfo {volume} {155}},\ \bibinfo {pages} {064702}
  (\bibinfo {year} {2021})}\BibitemShut {NoStop}%
\bibitem [{\citenamefont {Chen}\ \emph {et~al.}(2025)\citenamefont {Chen},
  \citenamefont {McKillop}, \citenamefont {Fidler},\ and\ \citenamefont
  {Weichman}}]{chen2025ultrafast}%
  \BibitemOpen
  \bibfield  {author} {\bibinfo {author} {\bibfnamefont {L.}~\bibnamefont
  {Chen}}, \bibinfo {author} {\bibfnamefont {A.~M.}\ \bibnamefont {McKillop}},
  \bibinfo {author} {\bibfnamefont {A.~P.}\ \bibnamefont {Fidler}},\ and\
  \bibinfo {author} {\bibfnamefont {M.~L.}\ \bibnamefont {Weichman}},\
  }\href@noop {} {\bibfield  {journal} {\bibinfo  {journal} {Nanophotonics}\
  }\textbf {\bibinfo {volume} {14}},\ \bibinfo {pages} {5437} (\bibinfo {year}
  {2025})}\BibitemShut {NoStop}%
\bibitem [{SM()}]{SM}%
  \BibitemOpen
  \href@noop {} {}\bibinfo {note} {See Supplemental Material at [URL will be
  inserted by publisher] for more on the treatment of intracavity fields and
  additional pump--probe simulations.}\BibitemShut {Stop}%
\bibitem [{\citenamefont {Ying}\ \emph {et~al.}(2025)\citenamefont {Ying},
  \citenamefont {Mondal}, \citenamefont {Koessler}, \citenamefont {Vega},\ and\
  \citenamefont {Huo}}]{ying2025collective}%
  \BibitemOpen
  \bibfield  {author} {\bibinfo {author} {\bibfnamefont {W.}~\bibnamefont
  {Ying}}, \bibinfo {author} {\bibfnamefont {M.~E.}\ \bibnamefont {Mondal}},
  \bibinfo {author} {\bibfnamefont {E.~R.}\ \bibnamefont {Koessler}}, \bibinfo
  {author} {\bibfnamefont {S.~M.}\ \bibnamefont {Vega}},\ and\ \bibinfo
  {author} {\bibfnamefont {P.}~\bibnamefont {Huo}},\ }\href@noop {} {\bibfield
  {journal} {\bibinfo  {journal} {Annu.\ Rev.\ Phys.\ Chem.}\ }\textbf
  {\bibinfo {volume} {77}} (\bibinfo {year} {2025})}\BibitemShut {NoStop}%
\end{thebibliography}%

\onecolumngrid
\newpage
\clearpage

	\renewcommand{\thesection}{S\Roman{section}}
	\renewcommand{\thesubsection}{S\Roman{section}.\Alph{subsection}}
	\renewcommand{\thetable}{S\arabic{table}}
	\renewcommand{\thefigure}{S\arabic{figure}}
	\renewcommand{\theequation}{S\arabic{equation}}
	\renewcommand{\thepage}{S\arabic{page}}
	
	\setcounter{section}{0}
	\setcounter{table}{0}
	\setcounter{figure}{0}
	\setcounter{page}{1}
	\setcounter{equation}{0}
		
	\centering 
 	\fontsize{14}{16}\selectfont	\textbf{Supplemental Material: Nonlinear Signal Enhancement of Strongly-Coupled Molecules in Pump--Probe Experiments} \\ \vskip1ex
	\fontsize{12}{13}\selectfont Alexander M. McKillop and Marissa L. Weichman\\
	\fontsize{11}{12}\selectfont \textit{ Department of Chemistry, Princeton University, Princeton, New Jersey 08544, USA}

	\justify	
	\fontsize{11}{11}\selectfont

		\section{SI. Representation of resonant and non-resonant pump and probe fields}
		
		To represent pump or probe fields that are resonantly injected into the cavity in the strongly-coupled region and therefore form standing waves, we employ the classical equation for the intracavity circulating field intensity [1]:
		\begin{align}
			\bigg| \tfrac{E_\textrm{RE}^\textrm{pu/pr}(\nu, z)}{E_0} \bigg|^2  
			= \tfrac{T \left[ e^{-\alpha(\nu)z} + R e^{-\alpha(\nu)(2L-z)} - 2\sqrt{R} e^{-\alpha(\nu)L} \cos \left[ \delta(\nu)(1-z/L) \right] \right]}{1 + R^2 e^{-2\alpha(\nu)L} - 2R e^{-\alpha(\nu)L} \cos \left[ \delta(\nu) \right] }
			\label{eq:Ere}
		\end{align}
		where $T$ and $R$ are the cavity mirror transmission and reflection intensity coefficients, $\alpha(\nu)$ is the absorption coefficient of the intracavity medium, and $L$ is the cavity length.
		$\delta(\nu)$ is the phase accrued by light making one round trip through the cavity, which takes on the value $2\pi m$ when light is injected resonantly into the $m^\textrm{th}$-order longitudinal cavity mode (or into polaritons formed from coupling this mode) [1].
		Here, we make the approximation that $\alpha(\nu)=0$ as absorption is typically small at the polariton frequencies,$^{29}$ and we assume lossless mirrors with $R=0.9$ and $T=0.1$.

		Meanwhile, non-resonant pump or probe fields pass through the cavity as traveling waves with decaying field intensity described using the Beer-Lambert Law: 
		\begin{align}
			\bigg| \tfrac{E_\textrm{NR}^\textrm{pu/pr}(\nu, z)}{E_0} \bigg|^2  =e^{-\alpha(\nu) z}
			\label{eq:Enr}
		\end{align}
		For non-resonant pumping of the $\textrm{S}_0\to \textrm{S}_2$ band in the main text, we assume a large ground state value of $\alpha(\nu)$ such that the field intensity is reduced by $50\%$ at $z=L$.
		For non-resonant probing of the $\textrm{S}_1\to \textrm{S}_\textrm{N}$ excited-state absorption in the main text, we take $\alpha(\nu) \approx 0$, assuming the transient absorption to be weak (typically on the scale of mOD in real experiments).

		\section{SII. Results with various single-molecule strong coupling threshold values}
		
		Signal enhancement from SC molecules occurs regardless of the choice of $g_\textrm{thresh}$ value, though the quantitative details do depend on this definition.
		%
		%
		To demonstrate how varying $g_\textrm{thresh}$ changes SC signal enhancement, we show the results for two cases below, with:
		%
		%
		%
		%
		\begin{align}
			g_\textrm{thresh,large} \equiv g_\textrm{max} \cdot \sin(\pi/3) \label{eq:S threshold large} \\[5pt]
			g_\textrm{thresh,small} \equiv g_\textrm{max} \cdot \sin(\pi/6) \label{eq:S threshold small}
		\end{align} 
		The definition of $g_\textrm{thresh,large}$ represents a more stringent threshold definition than that shown in the main text, where only molecules with single-molecule couplings larger than those of molecules at vacuum field anti-nodes which are oriented 60$^\textrm{o}$ from the $z$ axis are designated SC.
		The definition of $g_\textrm{thresh,small}$ is more relaxed than that shown in the main text, where molecules are designated SC if their single-molecule coupling exceeds that of a molecule at a vacuum field anti-node which is oriented 30$^\textrm{o}$ from the $z$ axis.

\begin{figure}[htbp]
			\centering \includegraphics[width=3in]{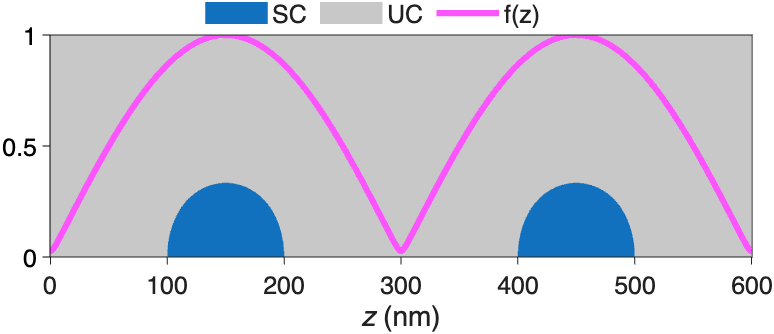}
			\caption{ \label{fig:S1} 
				Stacked bar graph of the fractions of strongly-coupled (SC, blue) and uncoupled (UC, gray) molecules along the $z$ axis of an $L=600$ nm cavity, considering resonant coupling to the $m=2$ longitudinal mode whose field profile is plotted in pink. 
				The molecules are assumed to be homogeneously distributed in position and orientation over 2000 $z$ points and 500 $\theta$ points.
				9\% of intracavity molecules meet the more stringent $g_\textrm{thresh,large}$ threshold for being SC defined in Eq.~\ref{eq:S threshold large}. 
			}
		\end{figure}

\begin{figure}[htbp]
			\centering \includegraphics[width=6in]{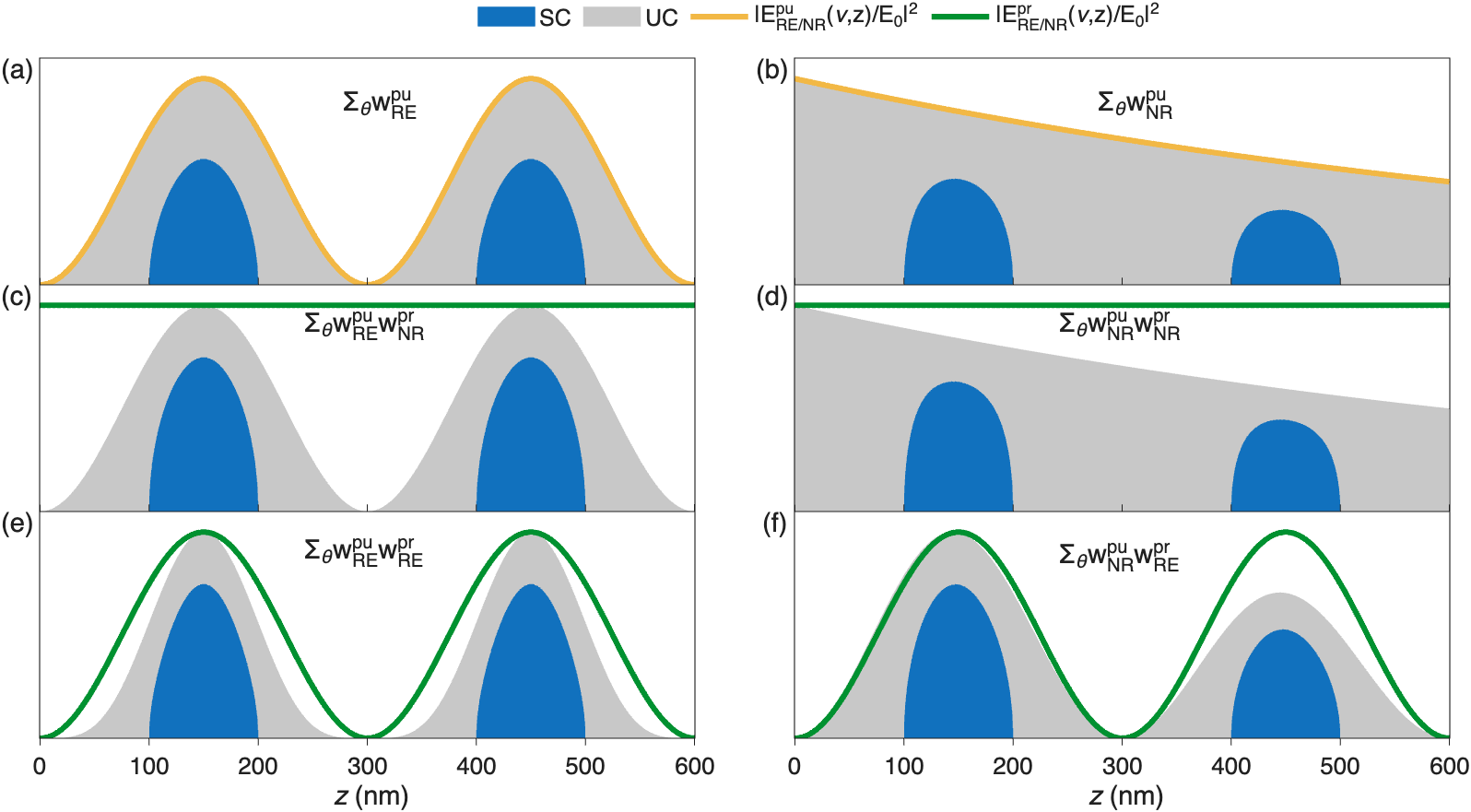}
			\caption{ \label{fig:S2} 
				Relative contribution of strongly coupled (SC, blue) and uncoupled (UC, grey) intracavity molecules to nonlinear pump--probe signals using the more stringent threshold in Eq.~\ref{eq:S threshold large} to determine which molecules are SC.
				(a,b) Stacked bar graphs showing transition weights for absorption of pump light, $w^\textrm{pu}$, by SC and UC molecules, summed over $\theta$ at each $z$ point for (a) RE and (b) NR pump light.
				Normalized RE and NR pump field intensities are superimposed as yellow lines. 
				(c--f) Stacked bar graphs showing the sum of transition weights for absorption of both pump and probe light, $w^\textrm{pu} w^\textrm{pr}$, by SC and UC molecules, summed over $\theta$ at each $z$ point for (c) RE--NR, (d) NR--NR, (e) RE--RE and (f) NR--RE pump-probe experiments.
				Normalized NR and RE probe field intensities are superimposed as green lines.
			}
		\end{figure}

		Figures \ref{fig:S1} and \ref{fig:S2} and Table~ \ref{tab:tableS1} show the results using $g_\textrm{thresh,large}$.
		As expected, the population of SC molecules is localized closely around the anti-nodes of the coupled cavity mode, and only $9\%$ of the total population is designated SC.
		%
		%
		Nonetheless, the enhancement factor of the nonlinear signals from SC molecules is greater with the $g_\textrm{thresh,large}$ threshold as compared to that used in the main text, as the more stringent definition of sC molecules also causes this population to interact more strongly with pump and probe fields.
		This is consistent with the conclusions drawn in the main text; that the same orientational factor which renders a molecule strongly-coupled also makes it contribute disproportionately to nonlinear signals.
		
\begin{figure}[htbp]
			\centering \includegraphics[width=3in]{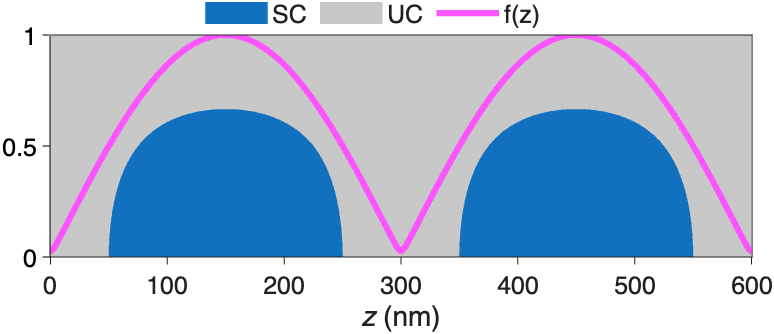}
			\caption{ \label{fig:S3} 
				Stacked bar graph of the fractions of strongly-coupled (SC, blue) and uncoupled (UC, gray) molecules along the $z$ axis of an $L=600$ nm cavity, considering resonant coupling to the $m=2$ longitudinal mode whose field profile is plotted in pink. 
				The molecules are assumed to be homogeneously distributed in position and orientation over 2000 $z$ points and 500 $\theta$ points.
				37\% of intracavity molecules meet the more stringent $g_\textrm{thresh,small}$ threshold for being SC defined in Eq.~\ref{eq:S threshold small}.  
			}
		\end{figure}

\begin{figure}[htbp]
			\centering \includegraphics[width=6in]{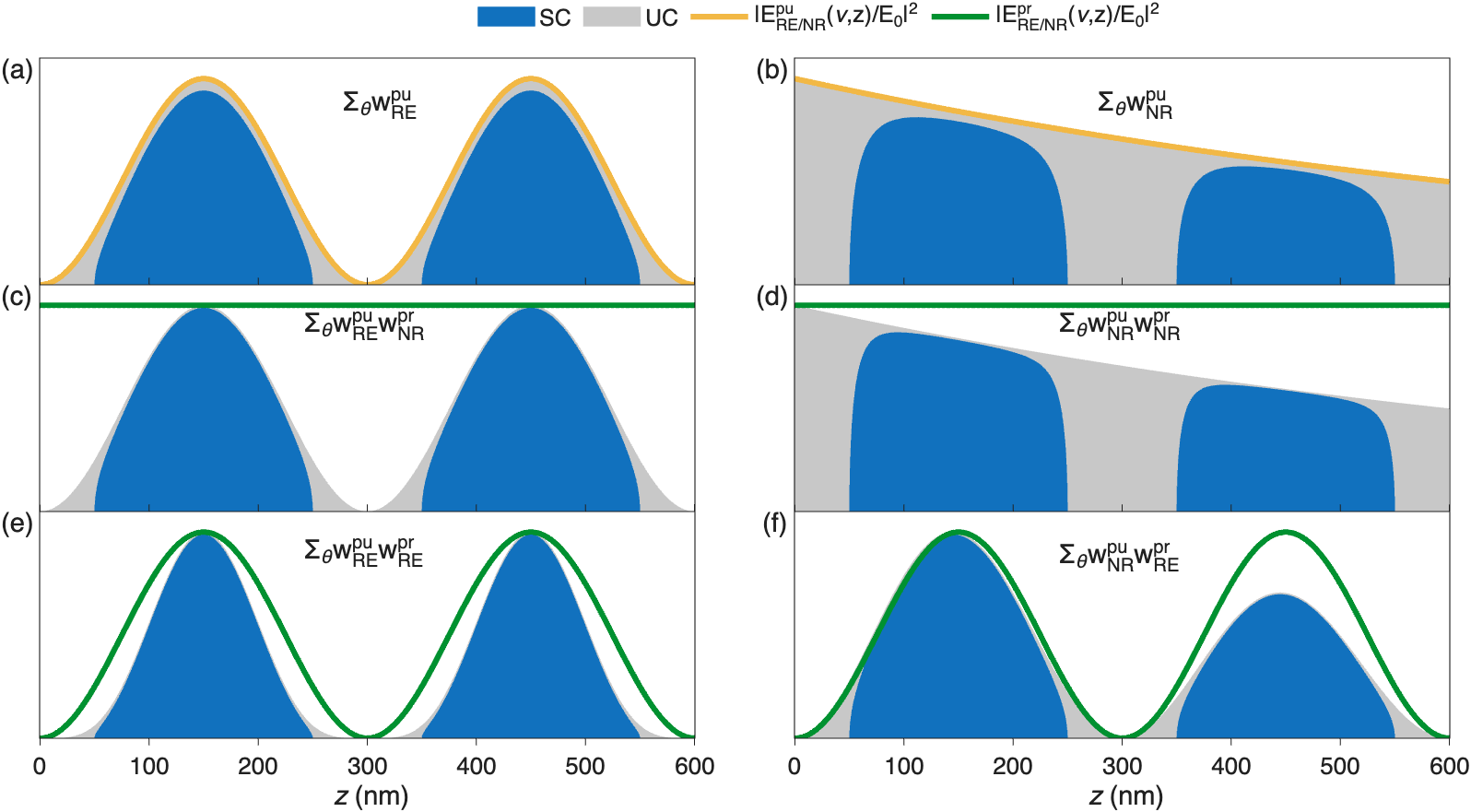}
			\caption{ \label{fig:S4} 
				Relative contribution of strongly coupled (SC, blue) and uncoupled (UC, grey) intracavity molecules to nonlinear pump--probe signals using the less stringent threshold in Eq.~\ref{eq:S threshold small} to determine which molecules are SC.
				(a,b) Stacked bar graphs showing transition weights for absorption of pump light, $w^\textrm{pu}$, by SC and UC molecules, summed over $\theta$ at each $z$ point for (a) RE and (b) NR pump light.
				Normalized RE and NR pump field intensities are superimposed as yellow lines. 
				(c--f) Stacked bar graphs showing the sum of transition weights for absorption of both pump and probe light, $w^\textrm{pu} w^\textrm{pr}$, by SC and UC molecules, summed over $\theta$ at each $z$ point for (c) RE--NR, (d) NR--NR, (e) RE--RE and (f) NR--RE pump-probe experiments.
				Normalized NR and RE probe field intensities are superimposed as green lines.
			}
		\end{figure}

		The results applying the $g_\textrm{thresh,small}$ threshold are summarized in Figs. \ref{fig:S3} and \ref{fig:S4} and Table~ \ref{tab:tableS2}. 
		Here, SC molecules make up $37\%$ of the intracavity population, but enhancement factors are lower.
		%
		
		%
		
		\noindent\makebox[\linewidth]{\rule{0.75\paperwidth}{0.4pt}}
		\noindent \small{ [1] A. M. McKillop and M. L. Weichman, Chem. Phys. Rev. \textbf{6}, 031308 (2025).}
		
		\pagebreak
		
		\begin{table}[H]
			\centering
			\caption{\label{tab:tableS1}
				Fractional contribution of SC molecules to pump--probe signals, and signal enhancement factors relative to the fraction of SC molecules ($9\%)$ set by $g_\textrm{thresh,large}$ in different experimental configurations. 
			}
			\begin{tblr}{ colspec={ c  c  c  c c }, rowsep = 3pt }
				\hline\hline
				& RE-NR & NR-NR & RE-RE & NR-RE \\ \hline
				$SC_\textrm{sig} (\%)$ & $38$ & $21$ &  $48$ & $38$ \\ \hline
				enhancement & $4.3$ & $2.3$ & $5.4$ & $4.3$ 
				\\ \hline\hline
			\end{tblr}
		\end{table}
		
		\pagebreak

		\begin{table}[htbp]
			\centering
			\caption{\label{tab:tableS2} 
				Fractional contribution of SC molecules to pump--probe signals, and signal enhancement factors relative to the fraction of SC molecules ($37\%)$ set by $g_\textrm{thresh,small}$ in different experimental configurations. 
			}
			\begin{tblr}{ colspec={ c  c  c  c c }, rowsep = 3pt }
				\hline\hline
				& RE-NR & NR-NR & RE-RE & NR-RE \\ \hline
				$SC_\textrm{sig} (\%)$ & $90$ & $61$ &  $96$ & $90$ \\ \hline
				enhancement & $2.4$ & $1.7$ & $2.6$ & $2.4$ 
				\\ \hline\hline
			\end{tblr}
		\end{table}

\end{document}